\pdfoutput=1

\documentclass[11pt,twoside,a4paper,cmspaper,final,collab]{cms-tdr}
\def\svnVersion{368093}\def\cmsCernNoTag{CERN-EP/2016-054}\def\cmsCernDate{\today}\def\cmsMessage{Published in the Journal of High Energy Physics as \href{http://dx.doi.org/10.1007/JHEP09(2016)051}{\doi{10.1007/JHEP09(2016)051.}}}

\begin{document}\cmsNoteHeader{HIG-14-032}

\hyphenation{had-ron-i-za-tion}
\hyphenation{cal-or-i-me-ter}
\hyphenation{de-vices}
\RCS$Revision: 364625 $
\RCS$HeadURL: svn+ssh://svn.cern.ch/reps/tdr2/papers/HIG-14-032/trunk/HIG-14-032.tex $
\RCS$Id: HIG-14-032.tex 364625 2016-08-15 20:52:32Z alverson $
\newlength\cmsFigWidth
\ifthenelse{\boolean{cms@external}}{\setlength\cmsFigWidth{0.85\columnwidth}}{\setlength\cmsFigWidth{0.4\textwidth}}
\ifthenelse{\boolean{cms@external}}{\providecommand{\cmsLeft}{top\xspace}}{\providecommand{\cmsLeft}{left\xspace}}
\ifthenelse{\boolean{cms@external}}{\providecommand{\cmsRight}{bottom\xspace}}{\providecommand{\cmsRight}{right\xspace}}
\newcommand{\mWW}{\ensuremath{m_{{\PW\PW}}}}
\newcommand{\GH}{\ensuremath{\Gamma_{\PH}}}
\newcommand{\GHs}{\ensuremath{\Gamma_{\PH}^{\mathrm{SM}}}}
\newcommand{\GHratio}{\ensuremath{\GH/\GHs}}
\newcommand{\HstrGF}{\ensuremath{\mu_\mathrm{GF}}}
\newcommand{\HstrVBF}{\ensuremath{\mu_\mathrm{VBF}}}
\newcommand{\HstrOffGF}{\ensuremath{\mu_\mathrm{GF}^{\text{off-shell}}}}
\newcommand{\HstrOffVBF}{\ensuremath{\mu_\mathrm{VBF}^{\text{off-shell}}}}
\newcommand{\HstrWwGF}{\ensuremath{\mu^{\PW\PW}_\mathrm{GF}}}
\newcommand{\HstrWwVBF}{\ensuremath{\mu^{\PW\PW}_\mathrm{VBF}}}
\newcommand{\HstrZzGF}{\ensuremath{\mu^{\cPZ\cPZ}_\mathrm{GF}}}
\newcommand{\HstrZzVBF}{\ensuremath{\mu^{\cPZ\cPZ}_\mathrm{VBF}}}
\newcommand{\met}{\ensuremath{\ET^{\text{miss}}}}
\newcommand{\mjj}{\ensuremath{m_{jj}}}
\newcommand{\dphill}{\ensuremath{\Delta\phi_{\Lep\Lep}}}
\newcommand{\mll}{\ensuremath{m_{\Lep\Lep}}}
\newcommand{\mth}{\ensuremath{m_{\mathrm{T}}^{\PH}}}
\newcommand{\WW}{\ensuremath{\PWp\PWm}\xspace}
\newcommand{\Hww}{\PH\to\WW}
\newcommand{\PV}{\ensuremath{\mathrm{V}}}
\newcommand{\dyll}{\ensuremath{{\cPZ/\Pgg^*\to \ell^+\ell^-}}}
\newcommand{\dytt}{\ensuremath{{{\cPZ}/\Pgg^* \to\TT}}}
\newcommand{\delR}{\ensuremath{\Delta R}}
\newcommand{\pmet}{\ensuremath{E^{\text{miss}}_{\mathrm{T},\text{min}}}}
\newcommand{\Lep}{\ensuremath{\ell}}
\cmsNoteHeader{HIG-14-032}
\title{Search for Higgs boson off-shell production in proton-proton collisions at 7 and 8\TeV
and derivation of constraints on its total decay width}

\date{\today}

\abstract{
  A search is presented for the Higgs boson off-shell production in gluon fusion and vector boson fusion processes with the Higgs boson decaying into a {$\WW$} pair and the {$\PW$} bosons decaying leptonically.
The data observed in this analysis are used to constrain the Higgs boson total decay width.
The analysis is based on the data collected by the CMS experiment at the LHC,
corresponding to integrated luminosities of 4.9\fbinv at a centre-of-mass energy of 7\TeV and 19.4\fbinv at 8\TeV, respectively.
An observed (expected) upper limit on the off-shell Higgs boson event yield normalised to the standard model prediction of 2.4 (6.2) is obtained
at the 95\% CL for the gluon fusion process
and of 19.3 (34.4) for the vector boson fusion process.
Observed and expected limits on the total width of 26 and 66\MeV are found, respectively, at the 95\% confidence level (CL).
These limits are combined with the previous result in the {$\cPZ\cPZ$} channel leading to observed and expected 95\% CL upper limits on the width of 13 and 26\MeV, respectively.
}

\hypersetup{%
pdfauthor={CMS Collaboration},%
pdftitle={Search for Higgs boson off-shell production in proton-proton collisions at 7 and 8 TeV
and derivation of constraints on its total decay width},%
pdfsubject={CMS},%
pdfkeywords={CMS, physics, Higgs, width, off-shell, WW, ZZ}}

\maketitle

\section{Introduction}
\label{sec:introduction}

A new particle, with properties consistent with those of the standard model (SM) Higgs boson $(\PH)$, was discovered at the CERN LHC with a mass near 125\GeV by the ATLAS and CMS collaborations~\cite{Aad:2012tfa, Chatrchyan:2012ufa, Chatrchyan:2013lba}.
Several properties of this particle have been measured to check its consistency with the SM~\cite{Chatrchyan:2012br,Aad:2013xqa,Aad:2013wqa,EPJC_PreciseMass7_8TeV,Chatrchyan:2013legacy,PRD92_spin_parity_HVVcoupling78TeV}.
Direct measurements of the total decay width of the Higgs boson {$(\GH)$} gave upper limits of 3.4\GeV in the $4\ell$
decay channel (where lepton, $\ell$, corresponds to either an electron or a muon)~\cite{Chatrchyan:2013legacy} and 2.4\GeV in the {$\Pgg\Pgg$} decay channel~\cite{gamgamLegacy_14,EPJC_PreciseMass7_8TeV}, which makes the particle compatible with a single narrow resonance.
At the LHC, the precision of direct width measurements is limited by
the instrumental resolution of the ATLAS and CMS experiments,
which is three orders of magnitude larger than the expected natural width for the SM Higgs boson, $\GHs \sim 4.1\MeV$~\cite{Heinemeyer:2013tqa}.
The ratio of the natural width of the discovered boson with respect to that of the SM Higgs boson was assessed by ATLAS~\cite{Aad:2015gba}
in the combination of all on-shell decay modes,
including invisible and undetectable ones,
and found to be $\GH / \GHs = 0.64^{+0.40}_{-0.25}$
under the model-dependent assumption that couplings of the 125\GeV boson to $\PW$ and $\cPZ$ bosons
could not be greater than those in the SM.
The sizable off-shell production of the Higgs boson can also be used to constrain its natural width.
A measurement of the relative off-shell and on-shell production provides direct information on $\GH$~\cite{Kauer:2012hd,Passarino:2012ri,Passarino:2014HiggsCat,Kauer:2013InadZero,CaolaMelnikov:1307.4935,CampbellEllisWilliams:1311.3589v1},
as long as the Higgs boson off- and on-shell production mechanisms are the same as in the SM
and the ratio of couplings governing off- and on-shell production remains unchanged with respect to the SM predictions.
For example, we assume that the dominant production mechanism is gluon fusion (GF) and not quark-antiquark annihilation.
Also, we assume that GF production is
dominated by the top quark loop
and there are no beyond-SM particles significantly contributing
in the entire on/off-shell mass range probed by the analysis.
Finally, the relative rate of off-shell and on-shell production depends on the tensor structure
of the couplings for the discovered boson~\cite{Gainer:2014hha,Englert:2014aca}.
Possible contributions from anomalous couplings are not considered in this analysis.

The CMS experiment already used off-shell production to constrain $\GH$, using {$\PH\to\cPZ\cPZ$} decays to {$4\ell$} and {$2\ell 2\Pgn$} final states, and obtained observed (expected) upper limits of
$\GH < 22\,(33)$\MeV at the 95\% confidence level (CL)~\cite{CMS:2014ala}.
The {$4\ell$} analysis was later updated~\cite{CMS:2014lifeTimeWidth4l} to include some improvements and allow for studies of anomalous {$\PH\to\cPZ\cPZ$} couplings via their effect on the off-shell production.

Similarly, ATLAS presented a study in the {$\cPZ\cPZ$} and {$\PW\PW$} channels that constrained the observed (expected)
upper limit on the off-shell event yield normalised to the SM prediction (signal strength $\mu$) to the range of 5.1--8.6 (6.7--11.0).
The range is determined by varying the {$\cPg\cPg\to\PW\PW$} and {$\cPg\cPg\to\cPZ\cPZ$} background $K$ factor within the uncertainty of the higher-order QCD correction~\cite{ATLAS:2015widthZZWW}.
An observed (expected) upper 95\% CL limit of $\GH <23\,(33)\MeV$ was obtained, assuming the background $K$ factor is equal to the signal $K$ factor.

This paper presents an analysis to constrain {$\GH$} and the off-shell signal strength in the leptonic final states of the {$\PH\to\PW\PW$} decay, based on the method proposed in Ref.~\cite{Campbell:BoundingWidthHWW}.
Our analysis follows the same methodology as used in the $\cPZ\cPZ$ analysis mentioned above~\cite{CMS:2014ala}.
The {$\PW\PW$} channel has worse mass resolution than {$\cPZ\cPZ$}, which affects the width measurement.
However, the {$\PW\PW$} channel benefits from a significantly larger branching fraction and a lower threshold for off-shell {$\PH\to\PW\PW$} production~\cite{CampbellEllisWilliams:1311.3589v1}.
To maximize sensitivity, the results of this analysis are combined with those obtained in the {$\PH\to\cPZ\cPZ$} channel~\cite{CMS:2014ala,CMS:2014lifeTimeWidth4l}.

The $\PW\PW$ and $\cPZ\cPZ$ analyses are based on proton-proton ($\Pp\Pp$)~collision data collected by the CMS experiment at the LHC in 2011 and 2012, corresponding to integrated luminosities of 4.9\fbinv and 19.4\fbinv at the center-of-mass energies 7 and 8\TeV, respectively~\cite{lumiPAS2012Winter,lumiPAS2012Summer}.

The paper is organized as follows: after a brief description of the CMS detector in Section 2, event datasets and Monte Carlo (MC) simulation samples are presented in Section 3.
The object reconstruction and event selection are described in Sections 4 and 5, respectively.
These are followed by the analysis strategy in Section 6 and a description of systematic uncertainties in Section 7.
The individual results for the {$\PH\to\PW\PW$} channel and the combination of these results with those from the $\cPZ\cPZ$ channels are reported in Sections 8 and 9, and the summary is given in Section 10.

\section{The CMS detector}

The central feature of the CMS apparatus is a superconducting solenoid of 6\unit{m} internal diameter, providing a magnetic field of 3.8\unit{T}.
Within the solenoid volume there are a silicon pixel and strip tracker, a lead tungstate crystal electromagnetic calorimeter (ECAL),
and a brass and scintillator hadron calorimeter, each composed of a barrel and two endcap sections.
Forward calorimeters extend the pseudorapidity coverage provided by the barrel and endcap detectors up to $\abs{\eta} < 5$.
Muons are measured in gas-ionization detectors embedded in the steel flux-return yoke outside the solenoid.
The missing transverse momentum vector {$\ptvecmiss$} is defined as the projection on the plane perpendicular to the beams of the negative vector sum of the momenta of all reconstructed particles in an event. Its magnitude is referred to as {$\ETmiss$}.
A more detailed description of the CMS detector, together with a definition of the coordinate system and the relevant kinematic variables can be found in~\cite{CMSDETECTOR}.

\section{Event datasets and Monte Carlo simulation samples}
\label{sec:Samples}

The explicit final state used is the different-flavor dilepton final state {$\WW\to\Pepm\Pgn\Pgm^\mp\Pgn$}.
The same-flavor dilepton final states {$\WW\to\Pep\Pgn\Pem\Pgn/\Pgmp\Pgn\Pgmm\Pgn$} are not considered, as they are overwhelmed by background from the Drell--Yan {$\dyll$} production.

The events are triggered by requiring the presence of either one or two high-$\pt$ electrons or muons with tight lepton identification and isolation criteria and with $\abs{\eta} < 2.4\,(2.5)$ for muon (electron)~\cite{Chatrchyan:2012xi,Khachatryan:2015hwa}.
Triggers with a single lepton have electron (muon) {$\pt$} thresholds ranging from 17 to 27\,(24)\GeV.
The higher thresholds are used for data taking periods with higher instantaneous luminosity.
For the dilepton triggers, one lepton with $\pt > 17\GeV$ and another with $\pt > 8\GeV$ are required.
The average combined trigger efficiency for events
that pass the full event selection is 96\% as measured in independent datasets obtained using different triggers.

\begin{figure}[h!]
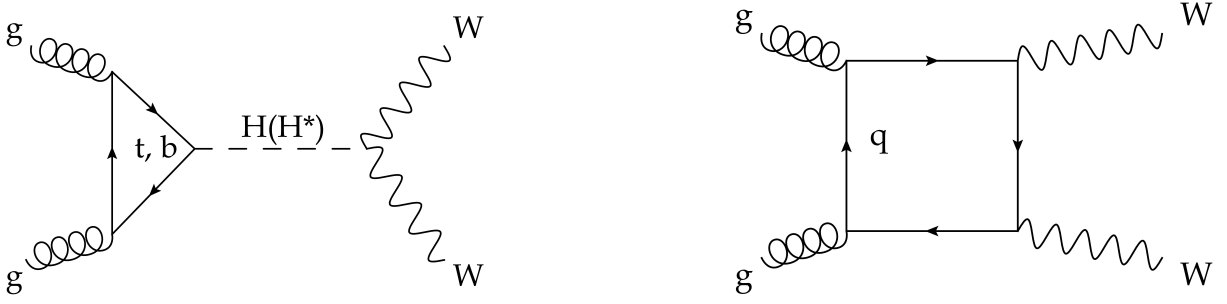

 \centering
    {\includegraphics[width=0.4\textwidth]{ggHWW.pdf}\label{fig:ggHWW}}\hfill
    {\includegraphics[width=0.4\textwidth]{gg_bkg.pdf}\label{fig:gg_bkg}}
    \caption{
      Feynman diagrams for the GF channel: (left) for the signal process {$\cPg\cPg\to\PH(\PH^{*})\to\PW^+\PW^-$},
      and (right) for the GF-initiated continuum background process {$\cPg\cPg\to\PW^+\PW^-$}.
	     The two processes can interfere, as they have identical initial and final states.}
    \label{fig:gg2WW}
\end{figure}

This analysis uses the dominant SM Higgs boson production modes of GF and vector boson fusion (VBF).
Other processes are not expected to contribute significantly to off-shell production~\cite{CMS:2014ala}.
The analysis accounts for possible interference between the Higgs boson signal and background processes when both have identical initial and final states.
Relevant leading order (LO) Feynman diagrams for GF and VBF processes for signal and background, which interfere with the signal, are depicted in Figs.~\ref{fig:gg2WW} and~\ref{fig:VBFWW}, respectively.
Following the previous study in the {$\cPZ\cPZ$} channels~\cite{CMS:2014ala},
a Higgs boson mass of {$m_{\PH} = 125.6\GeV$}~\cite{Chatrchyan:2013legacy}, with width {$\GH = 4.15\MeV$}~\cite{Heinemeyer:2013tqa},
is assumed for all of the event generation.
The small difference from the combined CMS and ATLAS Higgs boson mass, $125.1\pm 0.2\GeV$~\cite{CombinedHmassAtlasCms}, is found to have negligible impact on the width calculation.

\begin{figure}[h!]
 \centering
    {\includegraphics[width=0.33\textwidth]{qqHWW.pdf}\label{fig:qqHWW}}\hfill
    {\includegraphics[width=0.33\textwidth]{qq_bkg1.pdf}\label{fig:qq_bkg1}}\hfill
    {\includegraphics[width=0.33\textwidth]{qq_bkg2.pdf}\label{fig:qq_bkg2}}
    \caption{Feynman diagrams for the VBF channel: (left) for the signal process {$\Pq\Pq\to\Pq\Pq\PH(\PH^{*})\to\Pq\Pq\WW\to\Pq\Pq\ell^+\Pgn\ell^-\Pgn$},
      and (center and right) for two examples of background {$\Pq\Pq\to\Pq\Pq\WW\to\Pq\Pq\ell^+\Pgn\ell^-\Pgn$} channels.}
    \label{fig:VBFWW}
\end{figure}
The on-shell GF (VBF) signal, $\ttbar$, and {$\PQt\PW$} processes are generated with the {\POWHEG}~1.0 generator~\cite{powhegdiboson,Powheg_NLOQCD,POWHEG,NLOvectorPowheg,Powheg_POWHEGbOX}.
The other background processes, $\PW\cPZ$, $\cPZ\cPZ$, VVV (V = $\PW/\cPZ$), $\cPZ/\gamma^*$,
and $\qqbar\to \PW\PW$, are simulated using the \MADGRAPH 5.1 event generator~\cite{Alwall:2011uj}
as described in detail in the on-shell {$\Hww$} analysis~\cite{Chatrchyan:2013iaa}.

For the specific description of the Higgs boson off-shell region, the Higgs boson signal, the continuum {$\cPg\cPg\to\PW\PW$} background,
and their signal-background interference samples are generated using \textsc{gg2vv} 3.1.5~\cite{Binoth:2008pr} for GF production,
and \textsc{Phantom}~1.2.5~\cite{Ballestrero:2007} for VBF production at LO accuracy with the SM Higgs boson width.
The {CTEQ6L}~\cite{cteq66} LO parton distribution functions (PDF) are used by \textsc{gg2vv} and \textsc{Phantom}.
The dynamic factorization and renormalization scales of quantum chromodynamics (QCD) for \textsc{gg2vv} are
set to half the invariant mass of two $\PW$ bosons, {$\mu_F = \mu_R = \mWW/2$}.
For \textsc{Phantom} the QCD scale is set to
$Q^2 = M^2_{\PW} + \frac{1}{6}\sum^6_{i=1} p^2_{\mathrm{T}i}$, where $p_{\mathrm{T}i}$ denotes the transverse momentum of the $i$th particle in the final state with 6 particles defined in Fig.~\ref{fig:VBFWW}~\cite{Ballestrero:2007}.
The cross sections and various distributions at generator level obtained from \textsc{gg2vv} are
cross-checked by comparing them to {\MCFM}~6.8~\cite{MCFM} results.
For all processes,
the parton showering and the hadronization are implemented using {\PYTHIA} (version 6.422)~\cite{Pythia64}.

The {$K$}~factor for the GF process {$\cPg\cPg\to\PH\to\PW\PW$} is known up to next-to-next-to-leading-order (NNLO)~\cite{Bonvini:2013jha, SoftGluResumInterf}.
A value in the range 1.6--2.6 has been obtained with
an approximately flat dependency on $\mWW$.
For this analysis we use a value $K = 2.1$ affected by an uncertainty as large as 25\% as discussed in Section~\ref{sec:Systematics}.
A soft collinear approximation for the NNLO QCD calculation of the signal-background interference for the GF processes is reported in~\cite{Bonvini:2013jha},
which shows that the {$K$}~factor computed for the SM Higgs boson signal process is a good approximation to the interference process $K$~factor.
A similar study using soft gluon resummation confirms the same {$K$}~factor for the signal and the interference term at next-to-leading order (NLO) and NNLO~\cite{SoftGluResumInterf}.
The NLO QCD corrections to the LO background GF process, $\cPg\cPg\to\PW\PW$, are computed
in the heavy top quark approximation~\cite{ZZgluonHeavyTopApprox}, which shows that the {$K$}~factor for the background is similar to that for the signal.
Therefore, the {$K$}~factor calculated in the on-shell signal phase space is also used for the background and the interference term based on theoretical expectations.
The $K$~factor, defined as the ratio of NLO to LO cross sections for VBF production, has been shown to be close to unity by the NLO calculation of electroweak and QCD processes, with a 2\% theoretical uncertainty from missing higher-order effects~\cite{Ciccolini:2008EWkQCDcorrVBF}.
The QCD NNLO calculations~\cite{Bolzoni:2010xr,Bolzoni:2011cu} provide an identical cross section as obtained with the QCD NLO calculation
within a theoretical uncertainty of about 2\%.
Therefore the $K$~factor of the VBF process is set to unity with a 2\% theoretical uncertainty.

In the \textsc{gg2vv} samples, jets are generated by the parton shower algorithm implemented in \PYTHIA.
A better jet categorization is obtained with the NLO generator \POWHEG~1.0.
The jet multiplicity of the GF \textsc{gg2vv} sample is reweighted to take advantage of the jet description at the matrix element level in \POWHEG.
A "jet bin migration scale factor" is estimated as a function of the generator-level $\mWW$
by the comparison of the reconstruction-level \textsc{gg2vv} $\mWW$ spectrum to the \POWHEG $\mWW$ spectrum for each jet bin.
As an example, the jet bin migration scale factor for the 0-jet bin
varies by about 20\% in the range $160\GeV < \mWW < 1\TeV$, reducing the number of events in the 0-jet bin in the low-$\mWW$ region and increasing this number in the high-$\mWW$ region.
This jet bin migration scale factor is applied as a weight to the \textsc{gg2vv} sample used in this analysis.
The scale factor, calculated with the signal sample, is assumed to be the same for the background and interference samples.
The application of the factor to the background and interference samples has a negligible effect on the results.

The detector response is simulated using a detailed description of the CMS detector based on the \GEANTfour package~\cite{Agostinelli:2002hh}.  Minimum bias events are merged into the simulated events to reproduce the additional {$\Pp\Pp$} interactions in each bunch crossing (pileup).
The simulated samples are reweighted to represent the pileup distribution as measured in the data.
The average numbers of pileup interactions per beam crossing in the 7\TeV and 8\TeV data are about 9 and 21, respectively.

\section{Object reconstruction}
\label{sec:EventRec}

The particles candidates ($\Pe,\,\Pgm,$\,photon, charged hadron, and neutral hadron) in an event are reconstructed using the particle-flow algorithm~\cite{CMS-PAS-PFT-09-001,CMS-PAS-PFT-10-001}.
Clusters of energy deposition measured by the calorimeters, and tracks identified in the central tracking system and in the muon detectors, are combined to reconstruct individual particles.

Events used in this analysis are required to have two high-\pt lepton candidates (an electron and a muon) originating from a single primary vertex.
Among the vertices identified in an event, the one with the largest {$\sum \pt^2$},
where the sum runs over all tracks associated with the vertex, is selected as the primary vertex.

Electron candidates are defined by a reconstructed track in the tracking detector pointing to a cluster of energy deposition in the ECAL~\cite{Khachatryan:2015hwa}.
The electron energy is measured primarily from the ECAL cluster energy,
including bremsstrahlung recovery in the energy reconstruction by means of the standard CMS ECAL clustering algorithm.
A dedicated algorithm combines the momentum of the track and the ECAL cluster energy,
improving the energy resolution.
A multivariate approach is employed to identify electrons,
which combines several measured quantities describing track quality,
ECAL cluster shapes, and the compatibility of the measurements from the tracker and the ECAL.

A muon candidate is identified by the presence of a track in the muon system
matching a track reconstructed in the silicon tracker~\cite{Chatrchyan:2012xi}.
The precision of the measured momentum, based on the curvature of the track
in the magnetic field, is ensured by the acceptability criteria of the global fit
in the muon system and the hits in the silicon tracker.
Photon emission from a muon can affect the event reconstruction,
therefore a dedicated algorithm
identifies such cases and rejects the corresponding events.

Electrons and muons are required to be isolated to distinguish between prompt leptons
from {$\PW/\cPZ$} boson decays and leptons from hadron decays or misidentified leptons in multijet production.
Isolation criteria are based on the scalar sum of the transverse momenta of particles (scalar \pt sum) in the isolation cone defined by $\delR = \sqrt{\smash[b]{(\Delta\eta)^2 + (\Delta\phi)^2}}$ around the leptons.
The scalar \pt sum excludes the contribution of the candidate lepton itself.
To remove the contribution from the overlapping pileup interactions in this isolation region, the charged particles included in the computation of the isolation variable are required to originate from the primary vertex.
The contribution of pileup photons and neutral hadrons is estimated by the average particle \pt density deposited by neutral pileup particles, and is subtracted from the isolation cone~\cite{Cacciari:subtraction}.
The relative electron isolation is defined by the ratio of the scalar \pt sum in the isolation cone of {$\delR =~0.3$} to the transverse momentum of the candidate electron.
Isolated electrons are selected by requiring the relative isolation to be below {${\sim}$10\%}.
The exact threshold value depends on the electron {$\eta$} and {$\pt$}~\cite{XieThesis,10281_40097}.
For each muon candidate, the scalar \pt sum
is computed in
isolation cones of several radii around the muon direction.
This information is combined using a
multivariate algorithm that exploits the particles momentum deposition in the
isolation annuli to discriminate between prompt muons and the muons
from hadron decays inside a jet~\cite{Chatrchyan:2012xi}.

Jets are reconstructed using the anti-\kt clustering algorithm~\cite{Cacciari:2008gp}
with a distance parameter of 0.5, as implemented in the \FASTJET
package~\cite{Cacciari:fastjet1,Cacciari:fastjet2}.
A correction is applied to account for the pileup contribution to the jet energy similar to the correction applied for the lepton isolation.
A combinatorial background arises from low-$\pt$ jets from pileup interactions which get
clustered into high-$\pt$ jets.
A multivariate selection is adopted to separate jets from the
primary interaction and those reconstructed due to energy deposits associated with
pileup interactions~\cite{jetIdPAS}.
Jets considered for the event categorization are required
to have {$\pt > 30\GeV$} and {$\abs{\eta}<4.7\,(4.5)$} for the 8\,(7)\TeV analysis.

The identification of bottom {$(\PQb)$} quark decays is used to veto the background processes containing top quarks that subsequently decay to a {$\PQb$} quark and a {$\PW$} boson.
The {$\PQb$} quark decay is identified by {$\PQb$} quark jet ({$\PQb$} jet) tagging criteria based on the impact parameter significance of the constituent tracks or the presence of a soft muon in the event from the semileptonic decay of the {$\PQb$} quark~\cite{Chatrchyan:2012jua}.
For the former, the track counting high efficiency (TCHE) algorithm~\cite{TCHE:2010,Chatrchyan:2012jua} is used with a discriminator value greater than 2.1.
For the latter, soft muon candidates are defined without isolation requirements to be within $\delR=0.4$ from a jet and are required to have {$\pt > 3\GeV$}.
These {$\PQb$} tagging criteria retain {${\sim}95\%$} of the light-quark and gluon jets, while vetoing {${\sim}$70\%} of {$\PQb$} jets that arise from events with top quarks.

A projected~{$\met$} variable is defined as the component of {$\ptvecmiss$} transverse to the nearest lepton
if the lepton is situated within the {$\phi$} window of {$\pm \pi/2$} from the {$\ptvecmiss$} direction, otherwise  the projected~{$\met$} is the {$\met$} of the event.
A selection using this observable efficiently rejects {$\dytt$} background events, in which the {$\ptvecmiss$} is preferentially aligned with the leptons,
as well as {$\dyll$}
events with mismeasured {$\ptvecmiss$} caused by poorly reconstructed leptons.
Since the {$\ptvecmiss$} resolution is degraded by pileup, the minimum of two projected~{$\met$} variables is used {$(\pmet)$}: one constructed from all identified particles
(full projected {$\met$}), and another one from only the charged particles associated with the primary vertex (track projected {$\met$}).
The $\pmet$ has a better performance than either of the two correlated projected $\met$'s from which it is built as shown in Ref.~\cite{Chatrchyan:2013iaa}.

\section{Event selection}
\label{sec:EventSel}

Two main production processes are considered, GF and VBF, for which the method to determine {$\GH$} is identical, while event selections differ.
To increase the sensitivity to the SM Higgs boson signal,
events with a high-\pt lepton pair of different flavor (one electron and one muon, {$\Pe\Pgm$}) are selected,
and categorized according to jet multiplicities: zero jets (0-jet category), one jet (1-jet category), and two or more jets (2-jet category).
Higgs boson signal events in the 0- and 1-jet categories are mostly produced by the GF process, whereas the 2-jet category is more sensitive to the VBF production.

The $\PW\PW$ baseline selection criteria are the same as those used in the on-shell {$\PH\to\PW\PW$} analysis~\cite{Chatrchyan:2013iaa}.
For all jet multiplicity categories, candidate events are required to have two oppositely charged different-flavor leptons with {$\pt^{\ell_1} > 20\GeV$} for the leading lepton and {$\pt^{\ell_2} > 10\GeV$} for the sub-leading lepton.
Lepton pseudorapidities are restricted to be in the acceptance region of the detector, {$\abs{\eta}<2.5\,(2.4)$} for electrons (muons).
A small number of the electrons and muons considered in the analysis come from leptonic decays of {$\Pgt$} leptons after high \pt cuts of lepton.
Using simulation, the signal contribution of {$\Pgt$} leptonic decay from the {$\PH\to\PW\PW$} process,
with one or both {$\PW$} bosons decaying to {$\tau\nu$}, is estimated to be about 10\%.
The {$\pmet$} variable is required to be above 20\GeV to suppress $\dyll$ and $\dytt$ backgrounds.
The analysis requires the invariant mass of the dilepton $\mll > 12\GeV$
to reject the contributions from charmonium and bottomonium resonance decays.
Events having any {$\PQb$} jet are vetoed in order to suppress background events with top quarks.
The selection defined above is referred to as the {$\PW\PW$} baseline selection.

The GF selection consists of the {$\PW\PW$} baseline selection
and is applied to
events of the 0-jet and 1-jet categories.
The 2-jet category of the $\PW\PW$ baseline selection is enriched in VBF production
by requiring that
the two highest {$\pt$} jets are separated by {$\abs{\Delta\eta_{jj}} > 2.5$}.
In addition the pseudorapidity of each lepton {$i$} must obey the relation
{$\abs{\eta^{l_i} - (\eta^{j_1}+\eta^{j_2})/2}/\abs{\Delta\eta_{jj}}<0.5$}, where {$\eta^{l_i}$},
{$\eta^{j_1}$} and {$\eta^{j_2}$} are the pseudorapidities of the lepton and the two jets, and {$\Delta\eta_{jj}$} is the {$\eta$} distance between the two highest $\pt$ jets.
These cuts are based on the "VBF cuts" defined in Ref.~\cite{Zeppenfeld:1999HWwVBF}, exploiting the topology of VBF events.
The invariant mass {$\mjj$} of the two highest {$\pt$} jets must be larger than 500\GeV.
For events with three or more jets, the lowest {$\pt$} jets should not be between the two highest {$\pt$} jets in {$\eta$}.

\section{Analysis strategy}
\label{sec:AnalysisStrategy}

\begin{figure}[htbp]
 \captionsetup[subfigure]{margin=1pt}
 \centering
    \subfigure[GF 0-jet on-shell]{\includegraphics[width=0.42\textwidth]{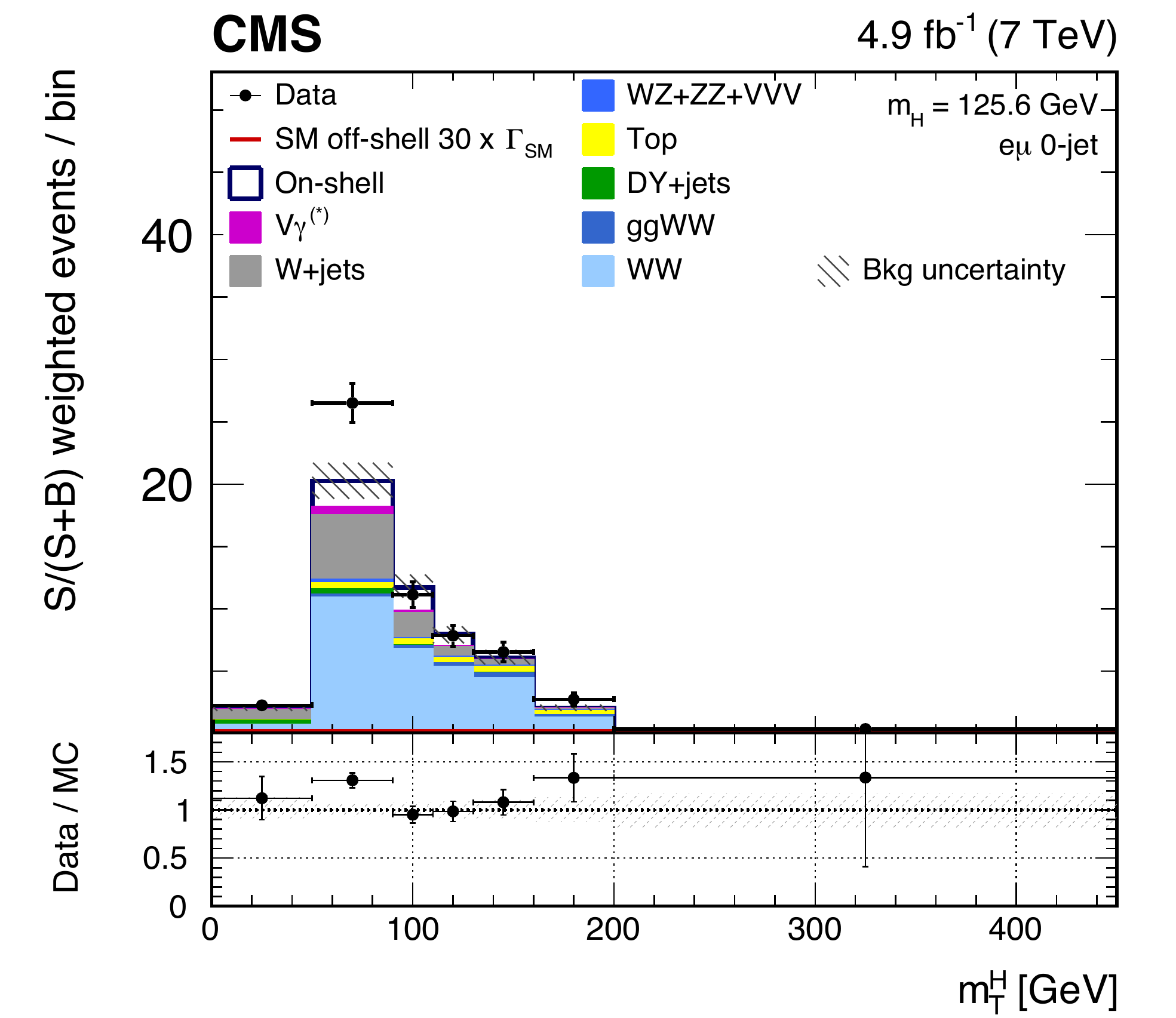}\label{fig:Histo1D_0jet_on_7TeV}}
    \subfigure[GF 0-jet off-shell]{\includegraphics[width=0.42\textwidth]{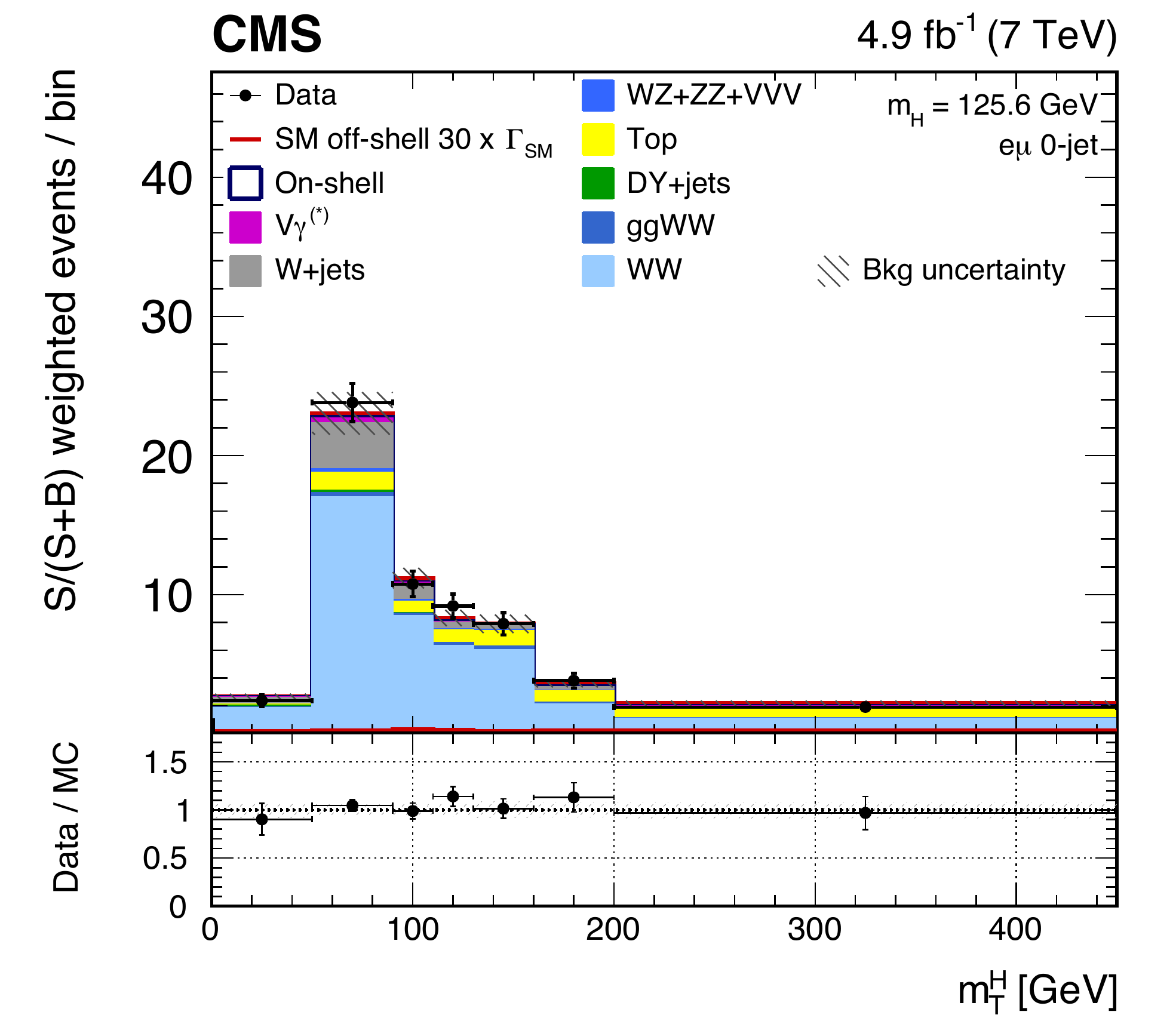}\label{fig:Histo1D_0jet_off_7TeV}}
    \\
    \subfigure[GF 1-jet on-shell]{\includegraphics[width=0.42\textwidth]{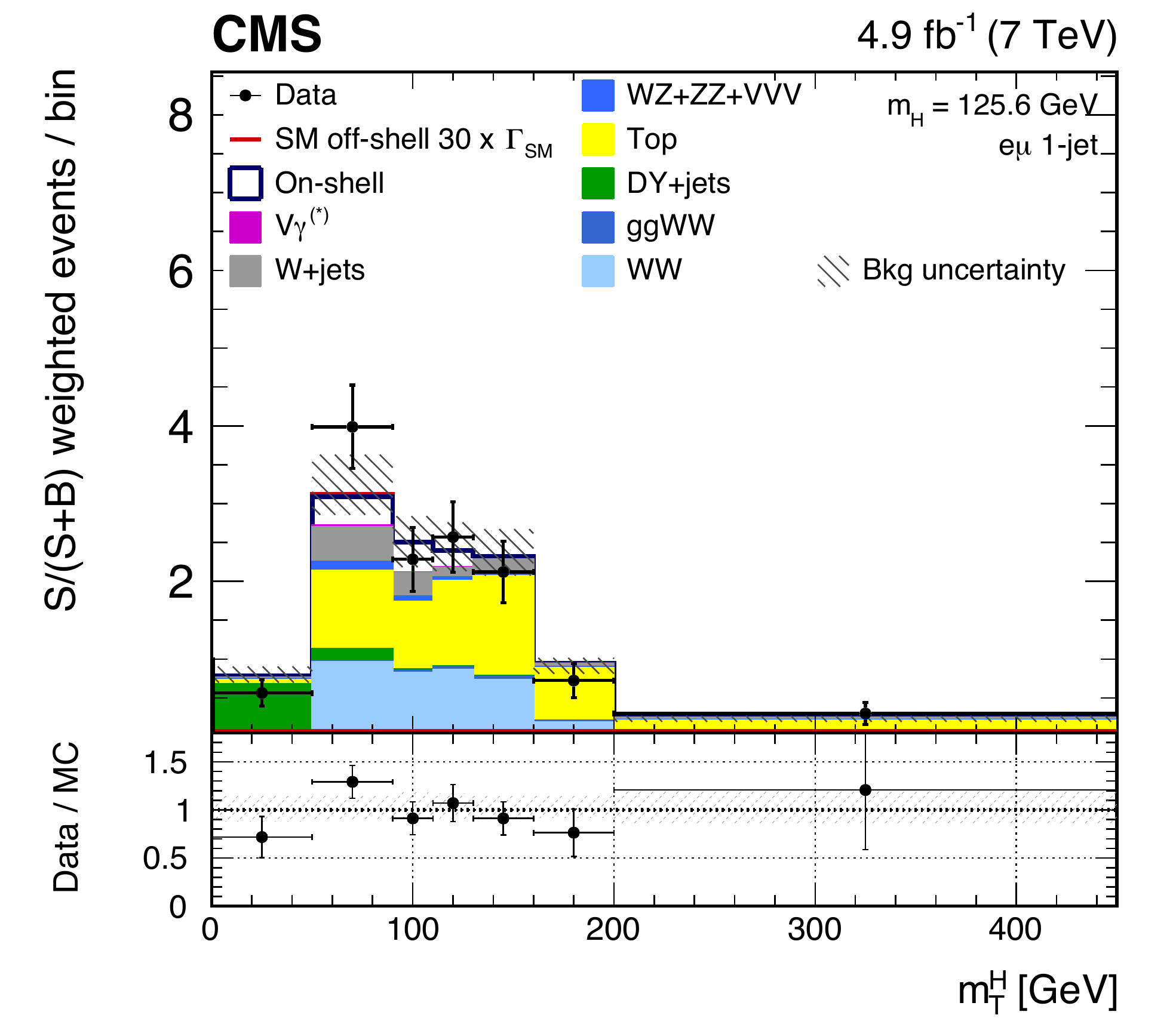}\label{fig:Histo1D_1jet_on_7TeV}}
    \subfigure[GF 1-jet off-shell]{\includegraphics[width=0.42\textwidth]{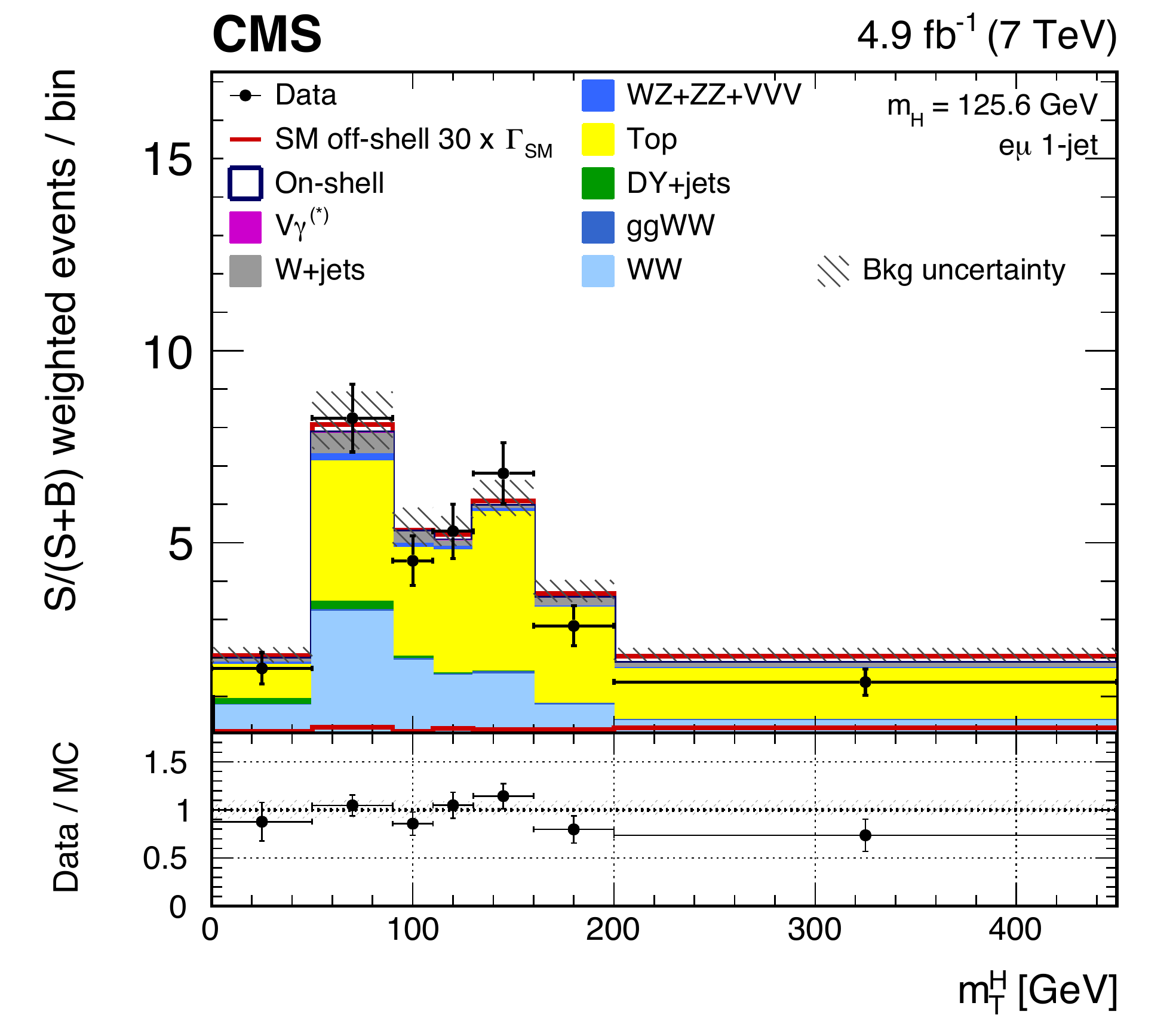}\label{fig:Histo1D_1jet_off_7TeV}}
    \\
    \subfigure[VBF 2-jet on-shell]{\includegraphics[width=0.42\textwidth]{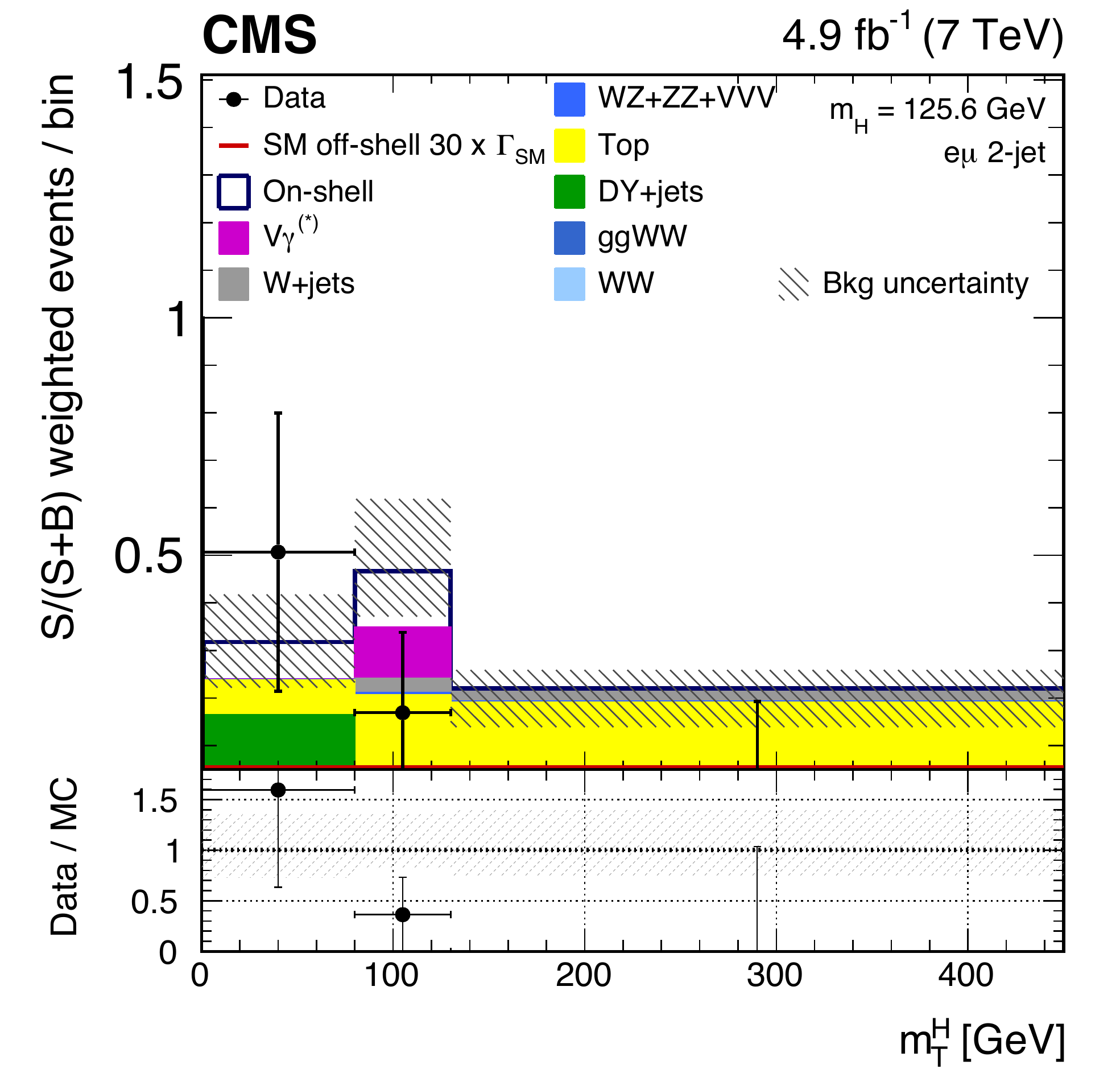}\label{fig:Histo1D_2jet_on_7TeV}}
    \subfigure[VBF 2-jet off-shell]{\includegraphics[width=0.42\textwidth]{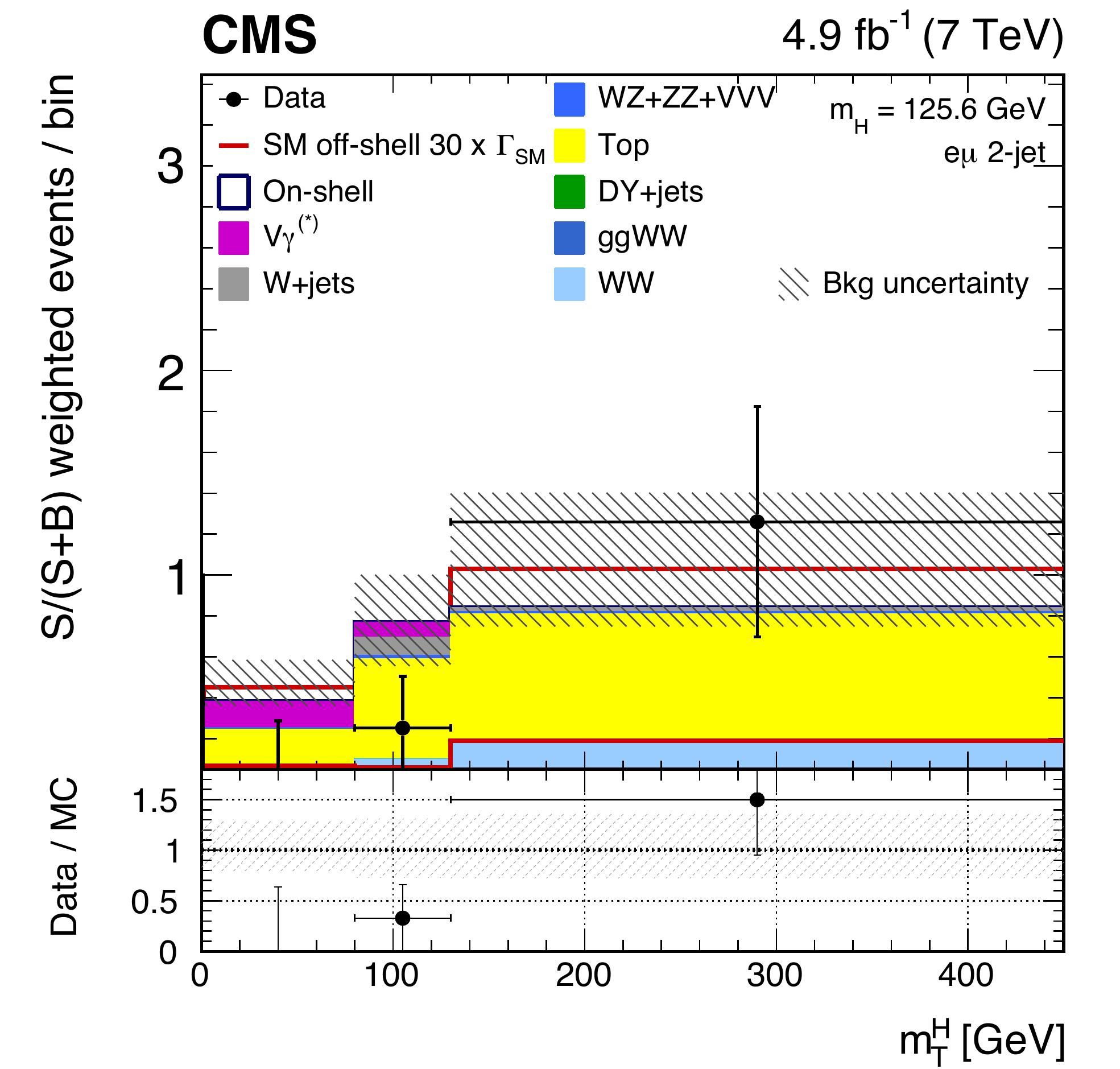}\label{fig:Histo1D_2jet_off_7TeV}}
    \caption{
             The {$\mth$} distributions for the GF 0-jet (a) and (b), and 1-jet (c) and (d) categories, and the VBF 2-jet category (e) and (f) for 7\TeV data.
	       The distributions are weighted as described in the text.
In the histogram panels, the expected off-shell SM Higgs boson signal rate, including signal-background interference,
   is calculated for $\GH=30\GHs$ and is shown with and without stacking on top of the backgrounds.
     In the data/MC panels, the expected off-shell SM Higgs boson rate is calculated for $\GH=\GHs$ for the comparison.
             }
    \label{fig:Histo01D_012jet_7TeV}
\end{figure}

\begin{figure}[htbp]
 \centering
    \subfigure[GF 0-jet on-shell]{\includegraphics[width=0.42\textwidth]{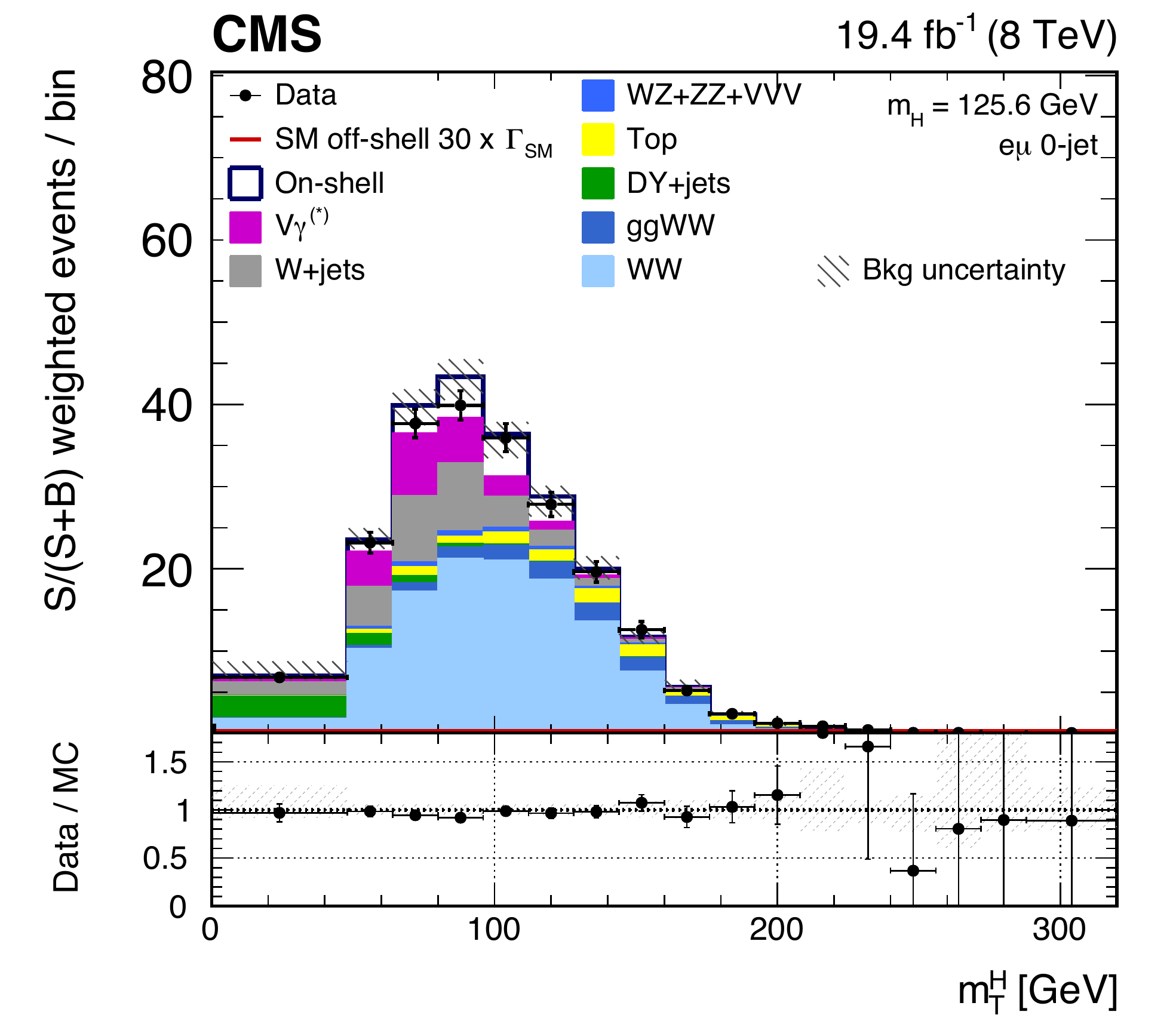}\label{fig:Histo1D_0jet_on}}
    \subfigure[GF 0-jet off-shell]{\includegraphics[width=0.42\textwidth]{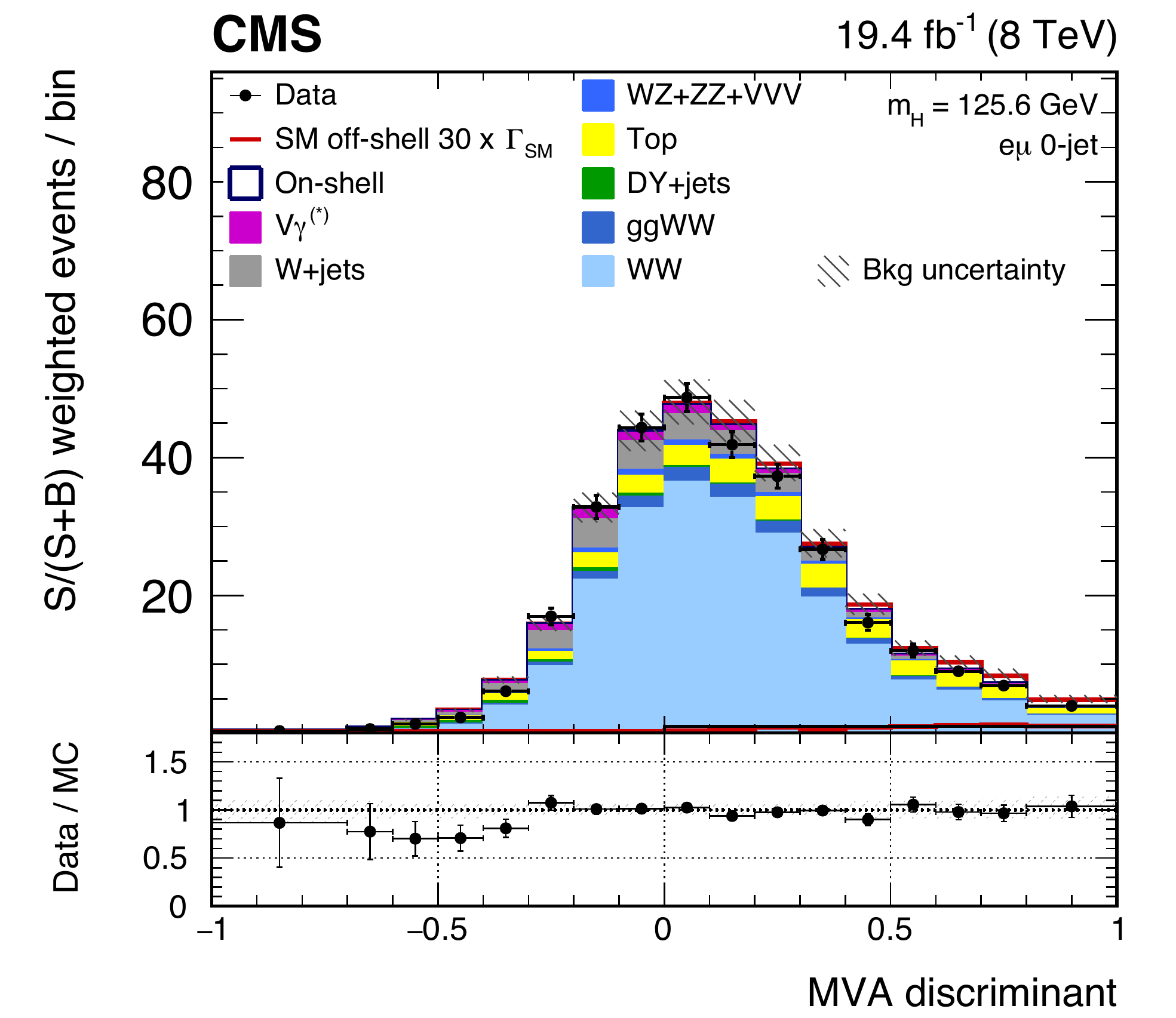}\label{fig:Histo1D_0jet_off}}
    \\
    \subfigure[GF 1-jet on-shell]{\includegraphics[width=0.42\textwidth]{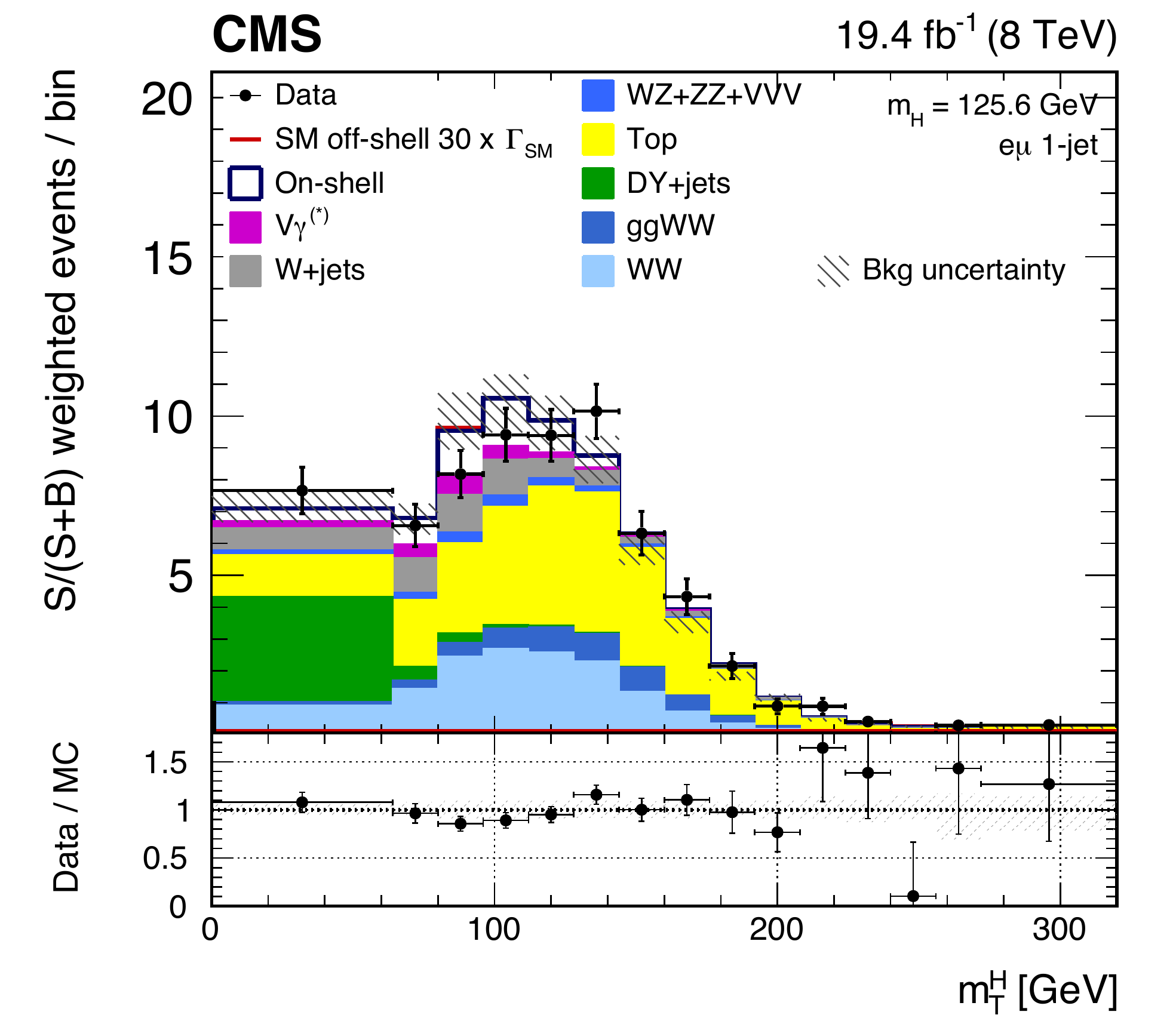}\label{fig:Histo1D_1jet_on}}
    \subfigure[GF 1-jet off-shell]{\includegraphics[width=0.42\textwidth]{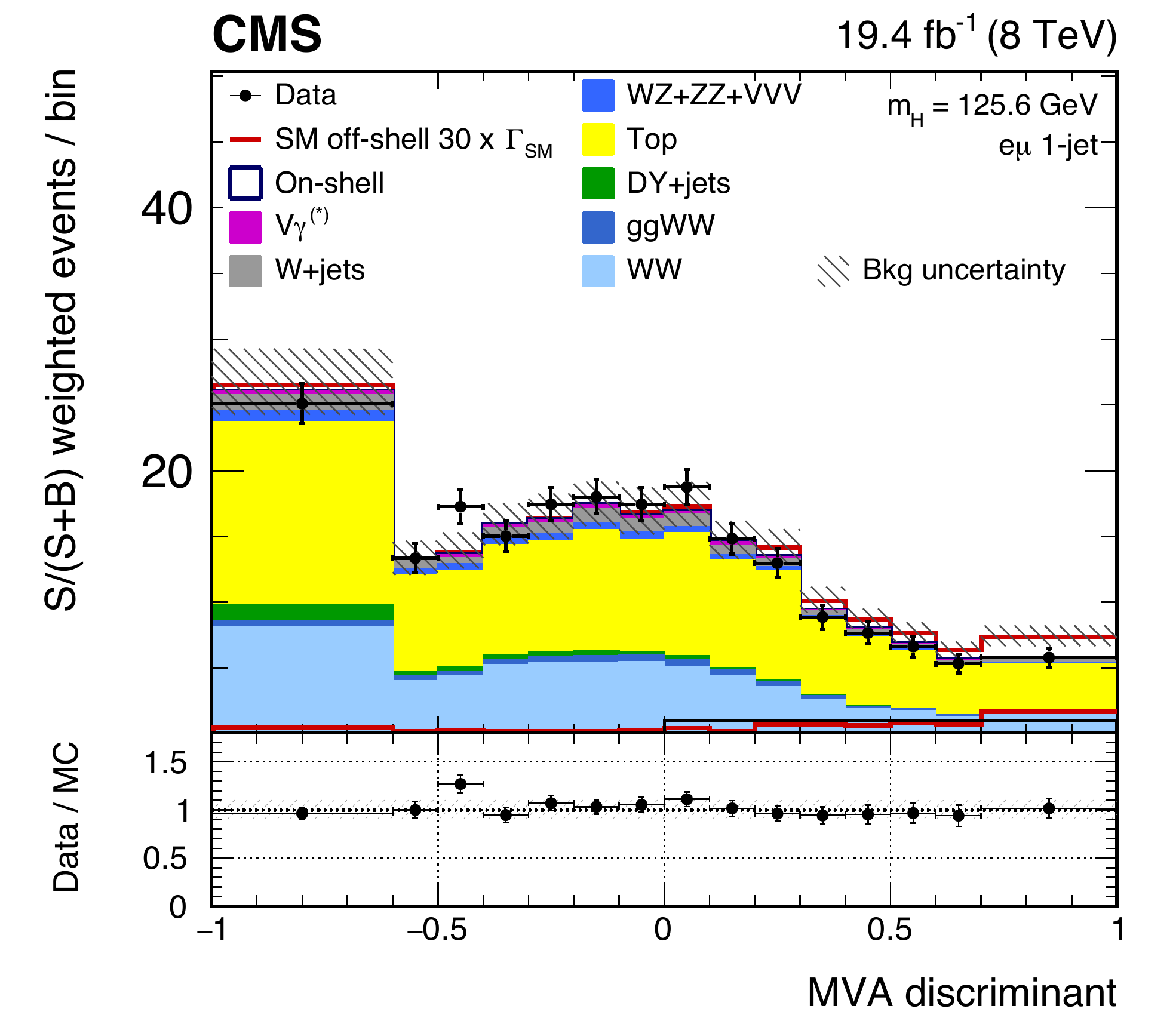}\label{fig:Histo1D_1jet_off}}
    \\
    \subfigure[VBF 2-jet on-shell]{\includegraphics[width=0.42\textwidth]{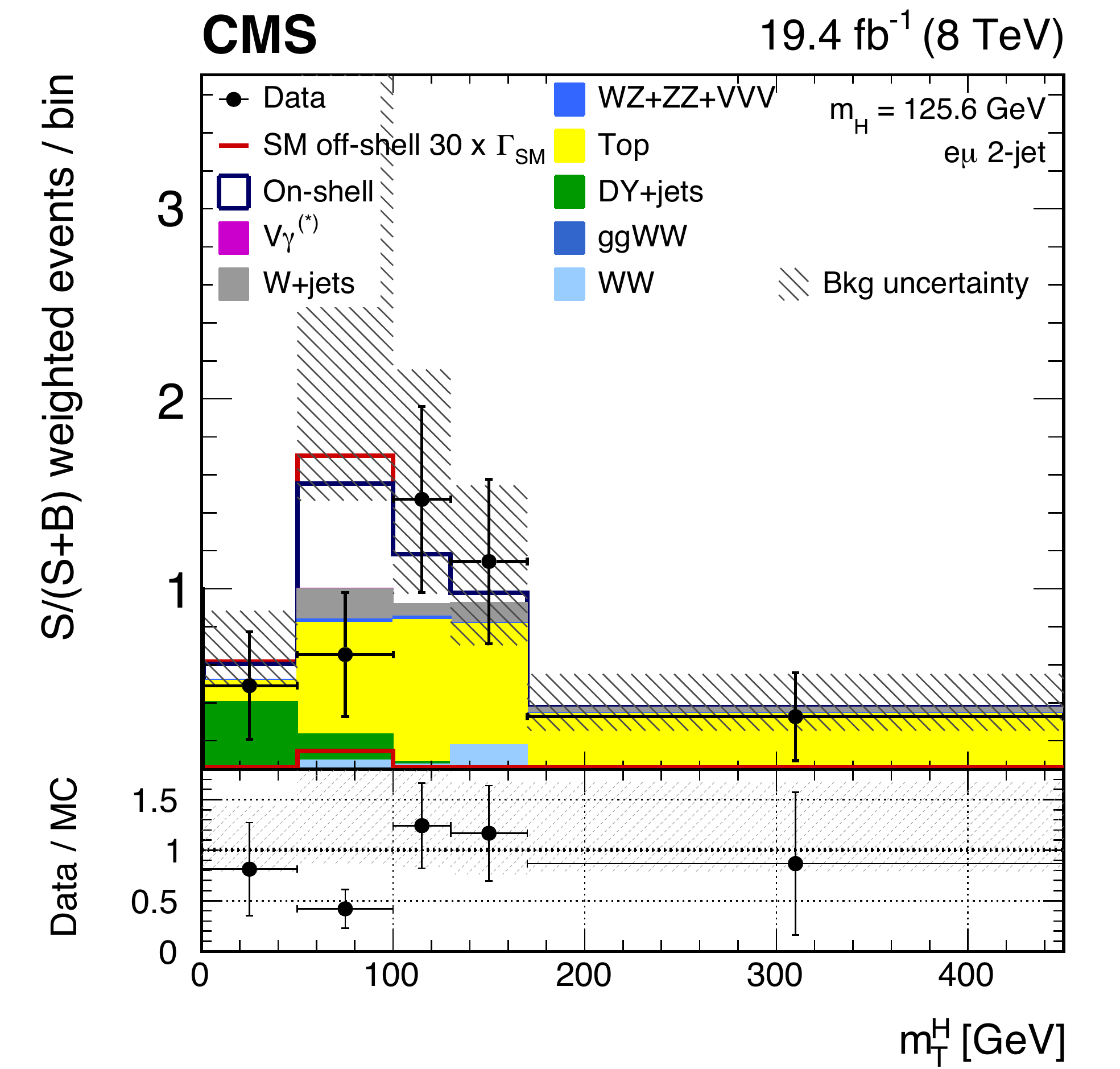}\label{fig:Histo1D_2jet_on}}
    \subfigure[VBF 2-jet off-shell]{\includegraphics[width=0.42\textwidth]{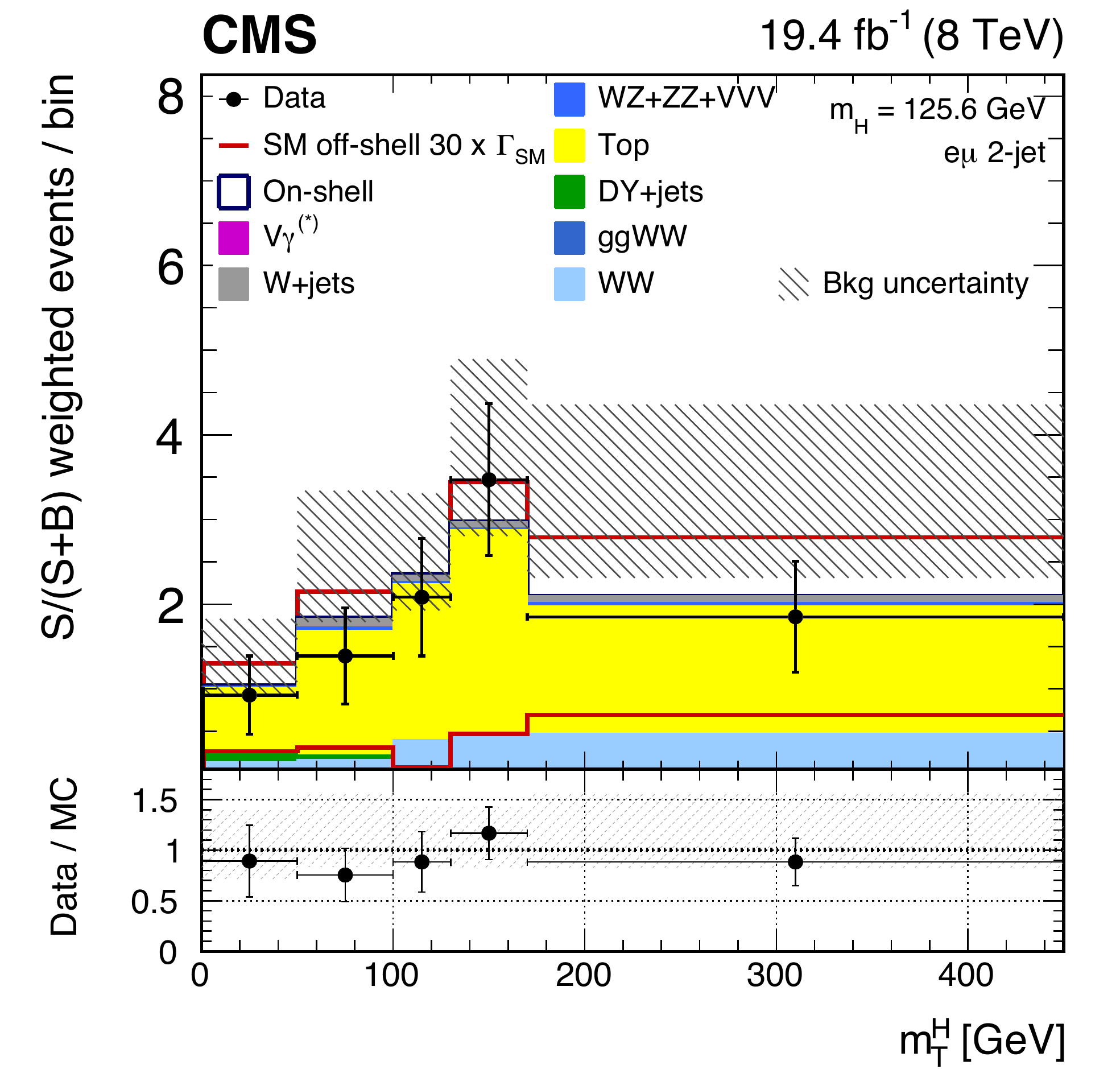}\label{fig:Histo1D_2jet_off}}
   \caption{
	    The {$\mth$} and MVA discriminant distributions for the GF 0-jet (a) and (b), and 1-jet (c) and (d) categories, and {$\mth$} for the VBF 2-jet category (e) and (f) for 8\TeV data.
	     More details are given in the caption of Fig.~\ref{fig:Histo01D_012jet_7TeV}.
             }
    \label{fig:Histo01D_012jet}
\end{figure}

The events retained after the {$\PW\PW$} baseline selection and the subsequent GF and VBF categorization
are further partitioned into two sub-samples.
The first sub-sample, where events are required to have {$\mll < 70\GeV$} is attributed to the on-shell Higgs boson category,
while the second sub-sample with {$\mll > 70\GeV$} is attributed to the off-shell Higgs boson category.
The expected on-shell Higgs boson signal is 196 (3) events in the on(off)-shell category
and the expected off-shell Higgs boson signal is 2 (7) events in the on(off)-shell category for 0-jet events after the baseline selection.
The level of on- and off-shell Higgs boson separation is shown in Figs.~\ref{fig:Histo01D_012jet_7TeV} and~\ref{fig:Histo01D_012jet} where the left (right) column shows the distributions in the on(off)-shell category.
The selection criteria for the on-shell category is the same as the previous on-shell {$\Hww$} study~\cite{Chatrchyan:2013iaa}, but is modified for the off-shell region as $\pt^{\ell\ell}> 45\GeV$ and ${\pt}^{\ell_2}> 20\GeV$
due to the different kinematics of signal and background production processes.
The transverse mass is defined as {$\mth = \sqrt{\smash[b]{ 2\pt^{\ell\ell} \met (1 - \cos \Delta \phi(\ptvec^{\ell\ell}, \ptvecmiss)) }}$},
where {$\ptvec^{\ell\ell}$} is the dilepton transverse momentum vector, {$\pt^{\ell\ell}$} is its magnitude, and {$\Delta \phi(\ptvec^{\ell\ell}, \ptvecmiss)$} is the azimuthal angle between the dilepton momentum and {$\ptvecmiss$}.
The {$\mth$} and the {$\mll$} are used to discriminate the Higgs boson signal from the dominant {$\PW\PW$} and top quark pair, {$\PW+\text{jets}$}, and {$\PW+\cPgg^{(*)}$} backgrounds.

\begin{figure}[h!]
 \centering
    \includegraphics[width=0.60\textwidth]{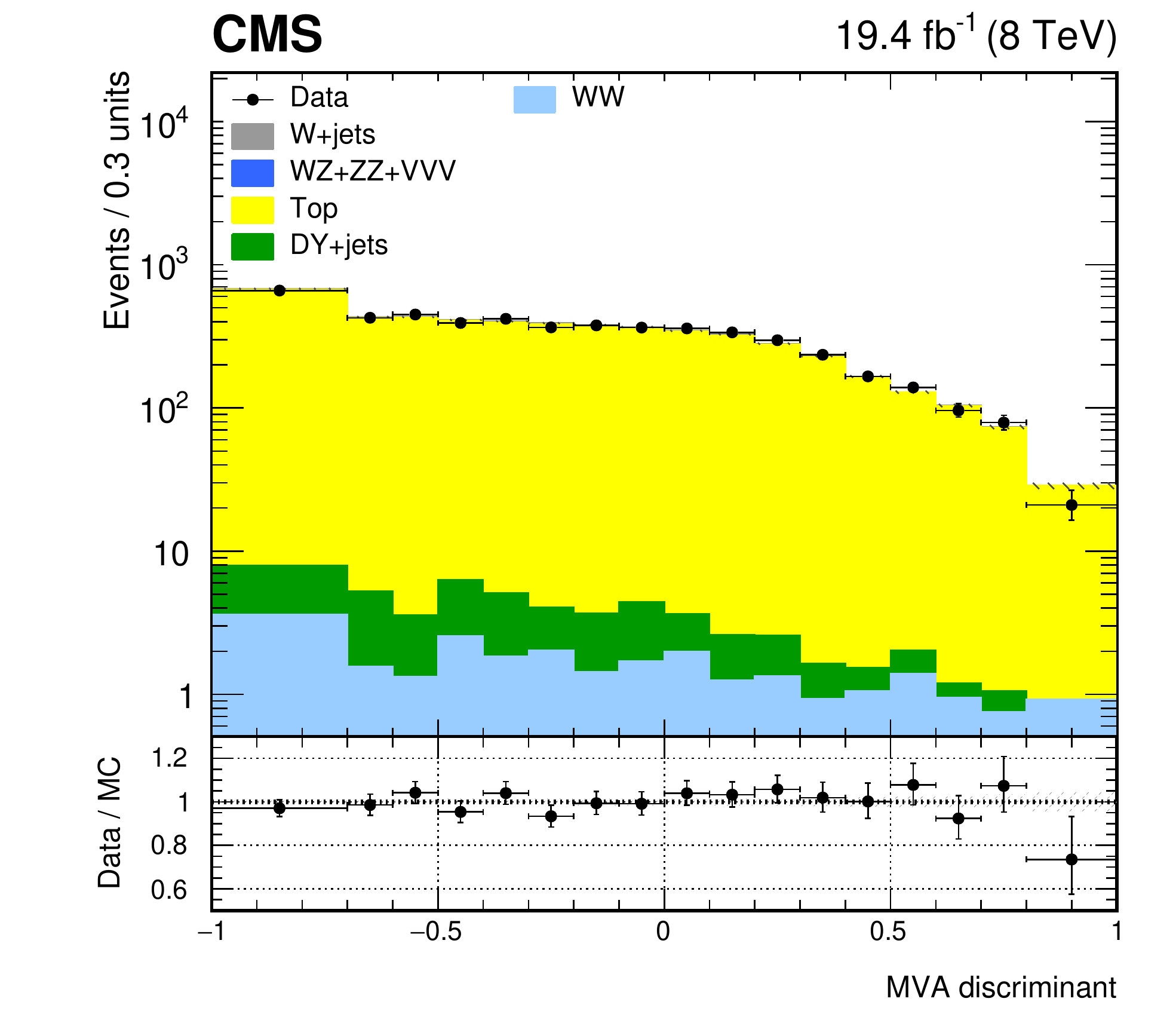}
    \caption{
             The MVA discriminant distribution for 8\TeV data for the 1-jet category in the top quark control region with one {$\PQb$}-tagged jet of
             {$\pt > 30\GeV$}. The {$\cPZ$}, {$\PW+\text{jets}$}, {$\PW\PW$}, and top quark simulation predictions are corrected with the estimates based on control samples in data, while other contributions are taken from simulation.
    }
    \label{fig:TopMVA}
\end{figure}

In order to enhance the sensitivity,
a boosted decision tree~\cite{BDT_Friedman} multivariate discriminator (MVA) is implemented with the toolkit for multivariate analysis (\textsc{tmva}) package~\cite{Hoecker2007:TMVA}
and is trained to discriminate between the off-shell Higgs boson signal and the other SM backgrounds.
Seven variables, {$\mth$}, {$\mll$}, the opening angle {$\dphill$} between the two leptons, {$\pt^{\ell\ell}$}, {$\MET$} in an event, {${\pt}^{\ell_1}$}, and {${\pt}^{\ell_2}$}, are used
for the boosted decision tree training and enter into the MVA discriminant.
Figure~\ref{fig:TopMVA} shows the MVA discriminant distribution tested on a top quark enriched region with 1 {$\PQb$}-tagged jet of {$\pt > 30\GeV$}, where good agreement between data and MC simulation is observed.
After validation of the MVA discriminant variable with 8\TeV MC simulation and data for the 0- and 1-jet categories,
the discrimination in these categories is performed using the {$\mll$} and MVA variables,
which achieve a 4\% improvement on the expected width limit compared to the {$\mll$} and {$\mth$} variables.
The analysis of other categories (8\TeV 2-jet category and all three of 7\TeV dataset categories)
use the {$\mll$} and {$\mth$} variables.
The selections and fit variables for the on and off-shell regions are given in Table~\ref{tab:RecoSelforLimits}.

\begin{table}[h]
\renewcommand{\arraystretch}{1.2}
  \centering
    \topcaption{Analysis region definitions for on- and off-shell selections.}
  \label{tab:RecoSelforLimits}
  \begin{tabular}{c|ccc}
  \hline
                     & On-shell            & Off-shell         & Off-shell\\
                     & (7, 8\TeV: all-jet) & (8\TeV: 0,1-jet)  & (7\TeV: all-jet, 8\TeV: 2-jet)\\ \cline{2-4}
    $\mll$           & $<$70\GeV        &  $>$70\GeV     & $>$70\GeV \\
  {$\pt^{\ell\ell}$} & $>$30\GeV        & $>$45\GeV      & $>$45\GeV\\
  {${\pt}^{\ell_2}$} & $>$10\GeV        & $>$20\GeV      & $>$20\GeV\\
  fit Var.           & $\mll$, $\mth$      & $\mll$, MVA       & $\mll$, $\mth$ \\
  \hline
  \end{tabular}
\end{table}

Twelve two-dimensional (2D) distributions {$\mll$} versus $\mth$ ({$\mll$} versus MVA for 8\TeV 0, 1-jet categories)
with variable bin size are defined.
The bin widths are optimized to achieve good separation between the SM Higgs boson signal and backgrounds,
while maintaining adequate statistical uncertainties in all the bins.
A 2D binned likelihood fit is performed simultaneously to these twelve distributions
using template 2D distributions which are obtained from the signal and background simulation.
For both the GF and VBF cases, expected event rates per bin are constructed
to be on-, or off-shell SM Higgs boson signal-like {$(\mathcal{P}_{\PH})$},
background-like {$(\mathcal{P}_{\text{bkg}})$} or interference-like {$(\mathcal{P}_{\text{int}})$}
defined in terms of the {$\mll$} and $\mth$ (MVA) observables.
To obtain a likelihood function depending on the SM Higgs boson GF (VBF) signal strength in the off-shell region $\HstrOffGF$ ($\HstrOffVBF$) without correlation to the on-shell GF (VBF) signal strength $\HstrGF$ ($\HstrVBF$),
the total expected event rates per bin $(\mathcal{P}_\text{tot}(\mll,\mth(\mathrm{MVA})|\text{$\mu$s}))$ can be written using these functions following~\cite{CaolaMelnikov:1307.4935, Anderson:2013afp} as
\begin{equation}\label{LikelihoodFtnStrength}\begin{split}
 \mathcal{P}_\text{tot}(\mll,\mth(\mathrm{MVA})|\text{$\mu$s}) =&
  \HstrOffGF \, \mathcal{P}^{\cPg\cPg}_{\PH,\,\text{off-shell}} +  \sqrt{\HstrOffGF} \, \mathcal{P}^{\cPg\cPg}_{\text{int}} + \mathcal{P}^{\cPg\cPg}_{\text{bkg}}
\\
 & +
  \HstrOffVBF \, \mathcal{P}^{\text{VBF}}_{\PH,\,\text{off-shell}} +  \sqrt{\HstrOffVBF} \,  \mathcal{P}^{\text{VBF}}_{\text{int}} + \mathcal{P}^{\text{VBF}}_{\text{bkg}}\\
& + \,
\HstrGF \, \mathcal{P}^{\cPg\cPg}_{\PH,\,\text{on-shell}}
+
\HstrVBF \, \mathcal{P}^{\text{VBF}}_{\PH,\,\text{on-shell}}
+
\mathcal{P}^{\cPq\cPaq}_{\text{bkg}} +
\mathcal{P}_{\text{other bkg}}.
\end{split}\end{equation}
Here,
{$\mathcal{P}^{\cPq\cPaq}_{\text{bkg}}$} is the contribution from the {$\cPq\cPaq\to\PW\PW$} continuum background, and
{$\mathcal{P}_{\text{other bkg}}$} includes the other background contributions.
Similarly, the likelihood function of the total width $\GH$ is obtained with the total expected event rates per bin $(\mathcal{P}_\text{tot}(\mll,\mth(\mathrm{MVA})|r))$
\begin{equation}\label{eq:pdf-prob-vbf}\begin{split}
 \mathcal{P}_\text{tot}(\mll,\mth(\mathrm{MVA})|r)  = &
  \HstrGF\, r \, \mathcal{P}^{\cPg\cPg}_{\PH,\,\text{off-shell}} +  \sqrt{\HstrGF\, r} \,  \mathcal{P}^{\cPg\cPg}_{\text{int}} + \mathcal{P}^{\cPg\cPg}_{\text{bkg}}
\\
& + \,
  \HstrVBF\, r \, \mathcal{P}^{\text{VBF}}_{\PH,\,\text{off-shell}} +  \sqrt{\HstrVBF\, r} \,  \mathcal{P}^{\text{VBF}}_{\text{int}} + \mathcal{P}^{\text{VBF}}_{\text{bkg}}
  \\
& + \,
\HstrGF \, \mathcal{P}^{\cPg\cPg}_{\PH,\,\text{on-shell}}
+
\HstrVBF \, \mathcal{P}^{\text{VBF}}_{\PH,\,\text{on-shell}}
+
\mathcal{P}^{\cPq\cPaq}_{\text{bkg}} +
\mathcal{P}_{\text{other bkg}},
\end{split}\end{equation}
where,
{$r=\GHratio$} is the scale factor with respect to the $\GHs$ determined by the Higgs boson mass value used in the simulation.

The normalisation and shape of the template 2D distributions used in the fit for the background processes
are obtained following the same procedure as in Ref.~\cite{Chatrchyan:2013iaa}.
Most of the background processes such as top quark, {$\PW\Pgg^*$}, and {$\PW+\text{jets}$} production, are estimated from data control regions.
The normalisation of the {$\cPq\cPaq\to\PW\PW$} background is constrained by the fit of $\mll$ versus $\mth$ or $\mll$ versus $\text{MVA}$ discriminant distribution using shapes determined by simulation.
For the 2-jet category, the $\PW\PW$ background normalization is taken from the MC simulation.
After the template fit to the {$\mll$} versus {$\mth$} (MVA) distributions for $\mu$s and $\GH$,
the observed projected {$\mth$} (MVA) distributions are compared to the fit results
in Figs.~\ref{fig:Histo01D_012jet_7TeV} and \ref{fig:Histo01D_012jet}.
In these figures, each process is normalized to the result of the 2D template fit and weighted using the other variable $\mll$.
This means that for the {$\mth$} (MVA) distributions,
the {$\mll$} distribution is used to compute the ratio of the fitted signal (S)
to the sum of signal and background (S+B) in each bin of the $\mll$ distribution integrated over the $\mth$ (MVA) variable.
In Fig.~\ref{fig:Histo01D_012jet_7TeV}, the observed {$\mth$} distributions are shown for
the GF mode 0- and 1-jet categories and for the VBF mode 2-jet category for 7\TeV data.
The {$\mth$} or MVA discriminant distributions of 8\TeV data are presented for the GF mode 0- and 1-jet categories and for the VBF mode 2-jet category in Fig.~\ref{fig:Histo01D_012jet}.

\section{Systematic uncertainties}
\label{sec:Systematics}

The systematic uncertainties for this analysis, presented in Table~\ref{tab:Systematics},
are classified into three categories as described in detail in Ref.~\cite{Chatrchyan:2013iaa}
and include uncertainties in the background yield predictions derived from data,
experimental uncertainties affecting normalisation and shapes of signal and backgrounds distributions obtained from simulation,
and theoretical uncertainties affecting signal and background yields estimated using simulation.

The dominant background for the 0-jet category is continuum {$\qqbar\to\PW\PW$} production.
The normalization of the {$\cPq\cPaq\to\PW\PW$} background for the 0 (1)-jet categories is determined from the 2D binned template fit to the data with 8\,(18)\% uncertainty dominated by the statistical uncertainty in the number of observed events.
The template 2D distribution obtained from the default generator is replaced by another one from \POWHEG to estimate the shape uncertainty in the fit.

Top quark production is the main background for the 1-jet and 2-jet categories.
Backgrounds from top quarks are identified and rejected via {$\PQb$} jet tagging based on the TCHE and the soft muon tagging algorithms.
The efficiency to identify top quark events is measured in a control sample dominated by {$\ttbar$} and {$\PQt\PW$} events, which is selected by requiring one {$\PQb$}-tagged jet.
The total uncertainty in the top quark background contribution is about 10\% for 0,1-jet and about 30\% for 2-jet category.
The scale of these uncertainties is defined by the control sample size (number of events) and the uncertainty of tagging algorithms.

The $\dytt$ background process is estimated using $\cPZ/\Pgg^*\to\mu\mu$ events selected in data,
in which muons are replaced with simulated $\tau$ decays.
The uncertainty in the estimation of this background process is about 10\%.

The non-prompt lepton background contributions originating from the leptonic decays of heavy quarks and {$\Pgt$} leptons, hadrons misidentified as leptons, and electrons
from photon conversions in {$\PW+\text{jets}$} and QCD multijet production, are suppressed by the identification and isolation requirements on electrons and muons, as described in Section~\ref{sec:EventRec}.
The remaining contribution from the non-prompt lepton background is estimated directly from data.
The efficiency, $\epsilon_\text{pass}$, for a jet that satisfies the loose lepton requirements
to pass the standard selection is determined using an independent sample dominated by events with non-prompt leptons from QCD multijet processes.
This efficiency is then used to weight the data with the loose selection
to obtain the estimated contribution from the non-prompt lepton background in the signal region.
The systematic uncertainty has two sources: the dependence of $\epsilon_\text{pass}$ on the sample composition, and the method.
The total uncertainty in $\epsilon_\text{pass}$, including the statistical precision of the control sample is about 40\%
for all cases (on- and off-shell, and all jet categories).

The contribution from {$\PW/\Pgg^{*}$} background processes is evaluated using a simulated sample, in which one lepton escapes detection.
The $K$ factor of the simulated sample is calculated by data control regions,
where a high-purity control sample of {$\PW/\Pgg^{*}$} events with three reconstructed lepton
is defined and compared to the simulation.
A factor of $1.5\pm0.5$ with respect to the LO prediction is found.
The shape of the discriminant variables used in the signal extraction for the {$\PW\Pgg$} process
is obtained from data control region that has 200 times more events than the simulated sample~\cite{Chatrchyan:2013iaa}.
The normalization is taken from simulated samples with uncertainty of 20\% dominated by the size of sample.

The integrated luminosity is measured using data from the HF system and the pixel detector~\cite{lumiPAS2012Winter,lumiPAS2012Summer}.
The uncertainties in the integrated luminosity measurement are 2.2\% at 7\TeV and 2.6\% at 8\TeV.

The lepton reconstruction efficiency in MC simulation is corrected to match data
using a control sample of {$\dyll$} events in the {$\cPZ$} boson peak
region~\cite{Khachatryan:2015hwa}.
The associated uncertainty is about 4\% for electrons and 3\% for muons.
The associated shape uncertainty is found to be negligible.

\begin{table}[h]
  \centering
  \topcaption{Summary of systematic uncertainties.}
  \label{tab:Systematics}
  \begin{tabular}{lc}
  \hline
  \multicolumn{2}{c} {Backgrounds estimated from data} \\
  \hline
  Source  & Uncertainty \\
  \hline
  {$\qqbar\to\PW\PW$}                & 8--18\% (0,1-jet) \\
  {$\ttbar$}, {$\PQt\PW$}   & {${\sim}10\%$} (0,1-jet); {${\sim}30\%$} (2-jet) \\
  {$\dytt$}        		& {${\sim}10\%$} \\
  {$\PW+\text{jet}$, QCD multijet}        & {${\sim}40\%$} \\
  {$\PW\Pgg/\Pgg^*$}        & {20--30\%} \\
  \hline
  \hline\\[-2.2ex]
  \multicolumn{2}{c} {Experimental uncertainties}\\
  \hline
  Source & Uncertainty\\
  \hline
  Integrated luminosity        & 2.2\% at 7\TeV 2.5\% at 8\TeV \\
  Lepton reconstruction and identification & 3--4\% \\
  Jet energy scale             & 10\% \\
  \hline \\[-2.2ex]
  \hline
  \multicolumn{2}{c} {Theoretical uncertainties}\\
  \hline
  Source & Uncertainty \\
  \hline
  {$\qqbar\to\PW\PW$}                & 20\% (2-jet) \\
  {$\PW\cPZ$}, {$\cPZ\cPZ$}, VVV & {${\sim}4\%$} \\
  QCD scale uncertainties: 		& \\
  \hspace{2em}On-shell signal                     & 20\% (GF); 2\% (VBF) \\
  \hspace{2em}Off-shell signal                    & 25\% (GF); 2\% (VBF) \\
  \hspace{2em}Bkg. and sig. + bkg. interf. 	      & 35\% (GF); 2\% (VBF) \\
  \hspace{2em}Exclusive jet bin fractions         & 30--50\% (GF); 3--11\% (VBF) \\
  PDFs                                 & 3--8\% \\
  Underlying event and parton shower   & 20\% (GF); 10\% (VBF) \\
  \hline
  \end{tabular}
\end{table}

Uncertainties in the jet energy scales affect the jet multiplicity and the jet kinematic variables.
The corresponding systematic uncertainties are computed by repeating the analysis with varied jet energy scales up and down by one standard deviation around their nominal values~\cite{Chatrchyan:2011ds}.
As a result, the uncertainty on the event selection efficiency is about 10\%.

For the 2-jet category, the {$\qqbar\to\PW\PW$} background rate is estimated from simulation with a theoretical uncertainty of 20\% by comparing two different generators \POWHEG and \MADGRAPH.

The total theoretical uncertainties in the diboson and multiboson production {$\PW\cPZ$}, {$\cPZ\cPZ$}, {$\PV\PV\PV$}, {$(\PV = \PW/\cPZ)$}, are estimated from the scale variation of renormalization and factorisation by a factor of two and are about 4\%~\cite{bib:ellis}.

The production cross sections and their uncertainties used for the SM Higgs boson expectation are taken
from Refs.~\cite{LHCHiggsCrossSectionWorkingGroup:2011ti,LHCHiggsCrossSectionWorkingGroup:2012ti}.
The uncertainties in the inclusive yields from missing higher-order corrections are evaluated by the change in the QCD factorization and renormalization scales and propagated to the $K$ factor uncertainty.
The $K$ factor uncertainty for the on-shell (off-shell) GF component is as large as 20 (25)\%
and it is 2\% for the VBF production in both on- and off-shell regions.
The $\cPg\cPg\to\PW\PW$ background and interference $K$ factors for GF production in the off-shell region are assumed to be the same as the signal $K$ factor with an additional 10\% uncertainty~\cite{Bonvini:2013jha,SoftGluResumInterf}.

The uncertainty on the predicted yield per jet bin associated with unknown higher order QCD corrections for GF are computed following
the Stewart--Tackmann procedure~\cite{StuTack}.
Samples have been produced with the {\sc sherpa}~2.1.1 generator~\cite{Gleisberg:2008ta, Cascioli:2011va, Cascioli:2013gfa},
which includes a jet at the QCD matrix element calculation for $\cPg\cPg\to\PW\PW$.
The factorization and renormalization scales are varied by factors of 1/2 and 2.
In the off-shell GF production, the uncertainty on the yield in each jet bin is about 30\% for the 0- and 1-jet cases and 50\% for the 2-jet case.
The effect of the large uncertainty in the 2-jet bin is negligible in the final results.

A similar comparison for the off-shell region is performed for the VBF process,
where the off-shell generation is provided by \textsc{Phantom}, which has LO accuracy.
Since two jets are generated at the matrix element level, the correction factor
to take into account jet bin migration is small and the uncertainty associated with it varies between 3\% and 11\%, depending on the jet bin.

The impact of variations in the choice of PDFs and
QCD coupling constant
on the yields is evaluated following the {\sc pdf4lhc} prescription~\cite{Alekhin:2011sk}, using the CT10, NNPDF2.1~\cite{nnpdf}, and MSTW2008 \cite{Martin:2009iq} PDF sets.
For the gluon-initiated signal processes (GF and {$\ttbar\PH$}), the PDF
uncertainty is about 8\%, while for the quark-initiated processes (VBF and Higgs boson production in association with a vector boson, {$\PV\PH$}) it is 3--5\%.

The systematic uncertainties due to the underlying event and parton shower model~\cite{ue_cms_7tev,ue_cms_7tev_dy} are
estimated by comparing samples simulated with different parton shower tunes and by disabling the underlying event simulation.
The uncertainties are around 20\% for GF and 10\% for VBF.

The overall sensitivity of the analysis to systematic uncertainties can be quantified as the relative difference in the observed limits on $\GH$ with and without systematic uncertainties included in the analysis;
it is found to be about 30\%.

\section{Constraints on Higgs boson width with \texorpdfstring{$\PW\PW$}{WW} decay mode}
\label{sec:Results_HWW}

Three separate likelihood scans are performed for the data observed in the twelve 2D distributions described in Section~\ref{sec:AnalysisStrategy}: $-2 \Delta \ln \mathcal{L}(\text{data}|\mu_{\mathrm{GF}}^{\text{off-shell}})$,
$-2 \Delta \ln \mathcal{L}(\text{data}|\mu_{\mathrm{VBF}}^{\text{off-shell}})$,
and  $-2\Delta\ln$$\mathcal{L}$(data$|\Gamma_{\mathrm{H}}$),
using data density functions defined by Eqs.~(\ref{LikelihoodFtnStrength}) and (\ref{eq:pdf-prob-vbf}),
where $-2 \Delta \ln \mathcal{L}$ is defined as
\begin{equation}\label{eq:likeliRatio}
-2 \Delta \ln \mathcal{L}( \text{data}|x ) = -2 \ln \frac{\mathcal{L}( \text{data}|x )}{\mathcal{L}_{\max}}.
\end{equation}
The profile likelihood function defined in Eq.~(\ref{eq:likeliRatio})
is assumed to follow a $\chi^2$ distribution (asymptotic approximation~\cite{AsymptoticFormulaeLHtest}).
We set 95\% CL limits on value $x$ from $-2 \Delta \ln \mathcal{L}( \text{data}|x ) = 3.84$.

\begin{figure}[htb]
 \centering
    \includegraphics[width=0.48\textwidth]{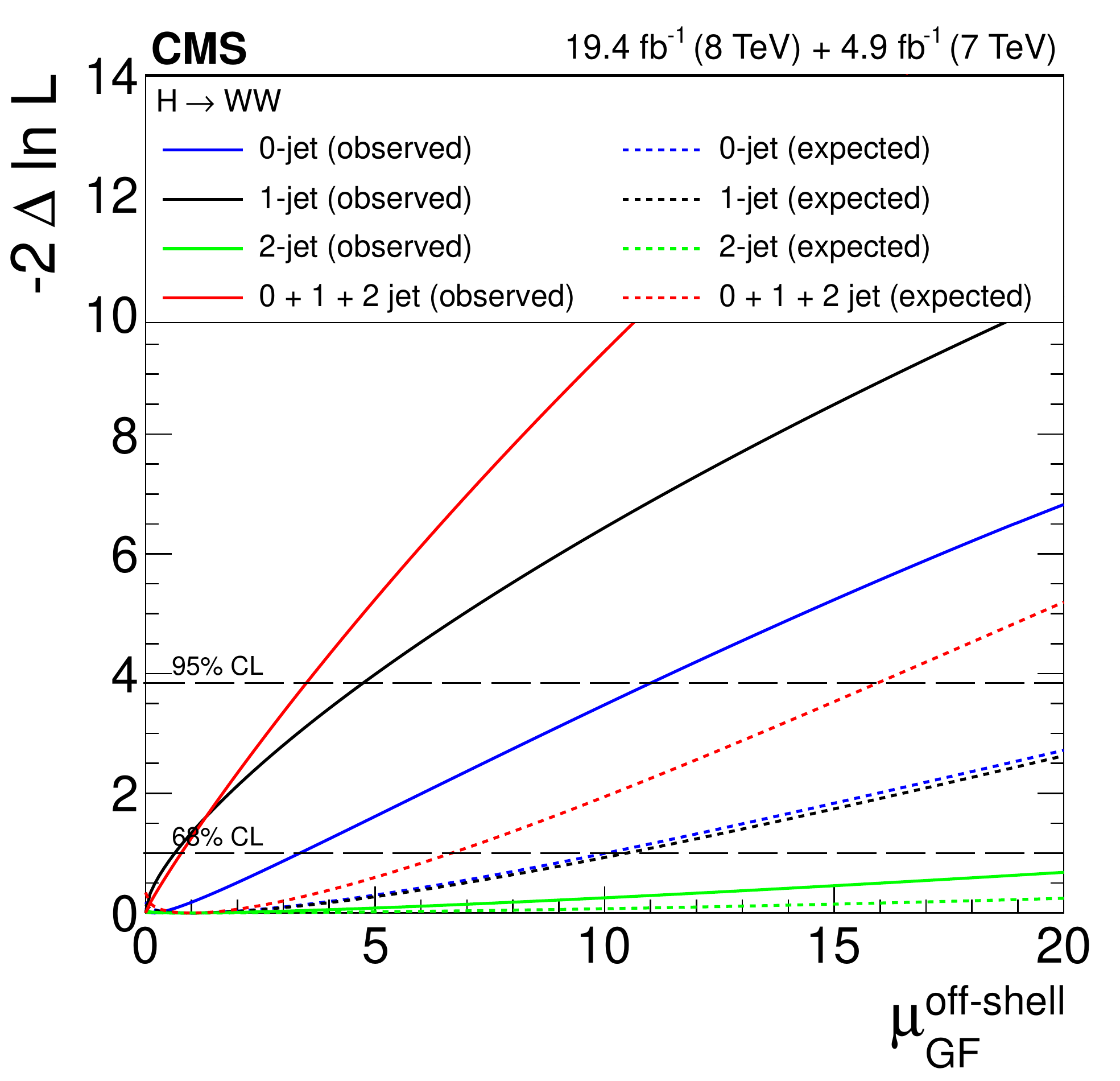}
    \includegraphics[width=0.48\textwidth]{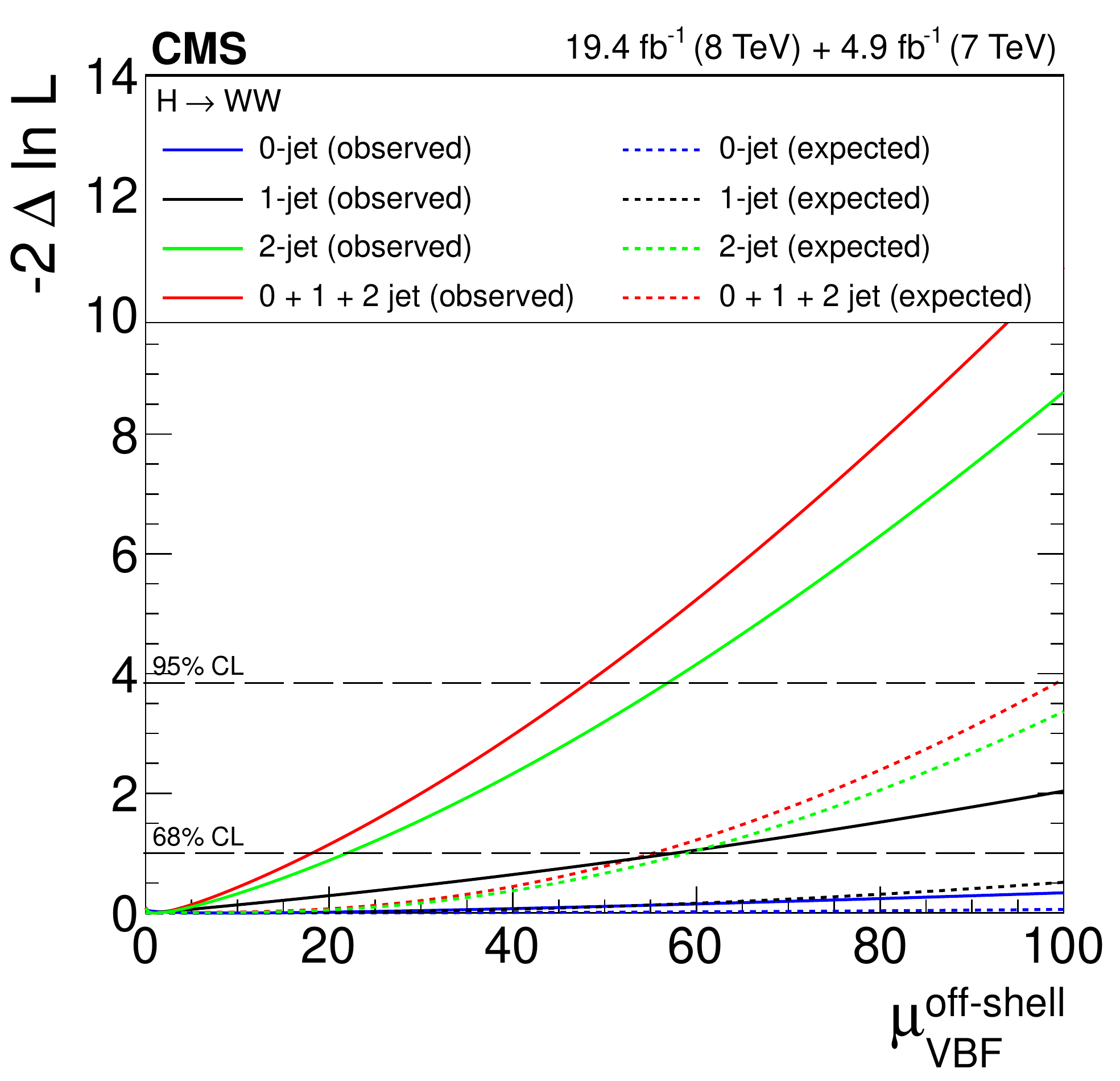}
    \caption{Scan of the negative log-likelihood as a function of the off-shell GF SM Higgs boson signal strength $\HstrOffGF$ (left)
	     and of the off-shell VBF signal strength $\HstrOffVBF$ (right)
	     for 0-, 1-, 2-jet categories separately and all categories combined for the {$\PH\to\PW\PW$} process:
             the observed (expected) scan is represented by the solid (dashed) line.
             }
    \label{fig:WWmuOffShell}
\end{figure}

When the negative log-likelihood, {$-2\Delta\ln {\cal L}$}, of $\HstrOffGF$ ($\HstrOffVBF$) is scanned, the other signal strengths are treated as nuisance parameters.
The uncertainties described in Section~\ref{sec:Systematics}
are incorporated as nuisance parameters in the scan.
The observed (expected) constraints of the off-shell signal strengths for six off-shell 2D distributions
(0-jet, 1-jet, 2-jet categories for 7 and 8\TeV data)
are $\HstrOffGF <3.5\,(16.0)$ and $\HstrOffVBF <48.1\,(99.2)$ at 95\% CL,
as shown in Fig.~\ref{fig:WWmuOffShell}.
The tighter than expected constraints arise from
the deficit in the observed number of events that is seen consistently in all jet categories
in the phase space most sensitive to the off-shell production, as shown in
Fig.~\ref{fig:Histo01D_012jet}.

\begin{figure}[htb]
 \centering
    \includegraphics[width=0.70\textwidth]{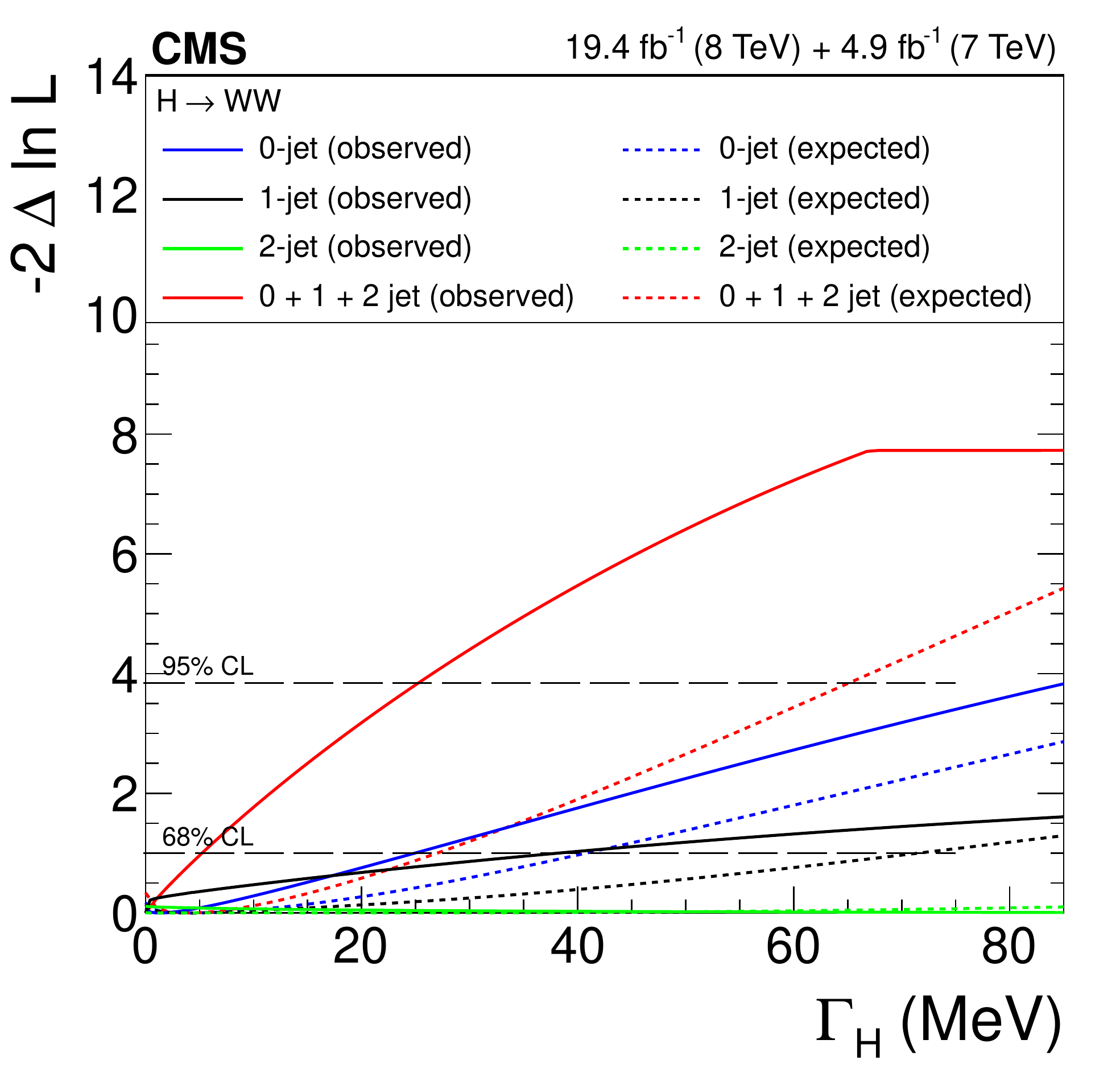}
    \caption{Scan of the negative log-likelihood as a function of {$\GH$}
	     for 0-, 1-, 2-jet categories separately and all categories combined for the {$\PH\to\PW\PW$} process:
             the observed (expected) scan is represented by the solid (dashed) line.
             }
    \label{fig:Limit_012jet}
\end{figure}

The results are shown in Fig.~\ref{fig:Limit_012jet} for scans of the likelihood as a function of {$\GH$}.
The {$\HstrGF$} and {$\HstrVBF$} are treated as nuisance parameters in the likelihood scan of {$\GH$}.
The scan combining the 0-, 1-, and 2-jet categories leads to an observed (expected) upper limit
of 26 (66)\MeV at 95\% CL on {$\GH$}.
Above $\GH=67\MeV$ the minimum value of $-2\Delta\ln{\cal L}$ stays constant at 7.7 corresponding to pure background hypothesis ($\HstrGF=0$, $\HstrVBF=0$): once the best-fit $\HstrGF$ and $\HstrVBF$ values reach zero, the likelihood given by Eq.~\ref{eq:pdf-prob-vbf} does not depend on $r$ anymore.

The coverage probability of the 95\% CL limit has been verified with toy MC simulation samples generated according to different $r$ hypotheses in Eq.~(\ref{eq:pdf-prob-vbf}).
The toy MC sample generated with $r=1$
has been used to estimate
the $p$-value of an observed limit of $<$26\MeV, while the expected one is $<66$\MeV.
A $p$-value of 3.6\% is obtained.

\section{Constraints on Higgs width with WW and ZZ decay modes}
\label{sec:Results_HWWandHZZ}

To exploit the full power of the Higgs boson width measurement technique based on the off-shell Higgs boson production approach,
the results using {$\PH\to\PW\PW$} reported here are combined with those found using {$\PH\to\cPZ\cPZ$}~\cite{CMS:2014ala,CMS:2014lifeTimeWidth4l}.
The {$\PH\to\cPZ\cPZ$} results are obtained using datasets corresponding to an integrated luminosity of 5.1 (19.7)\fbinv at 7 (8)\TeV.
The statistical methodology used in this combination is the same as the one employed in Ref.~\cite{CMS:2014ala}.

\begin{figure*}
\centering
     \includegraphics[width=0.48\textwidth]{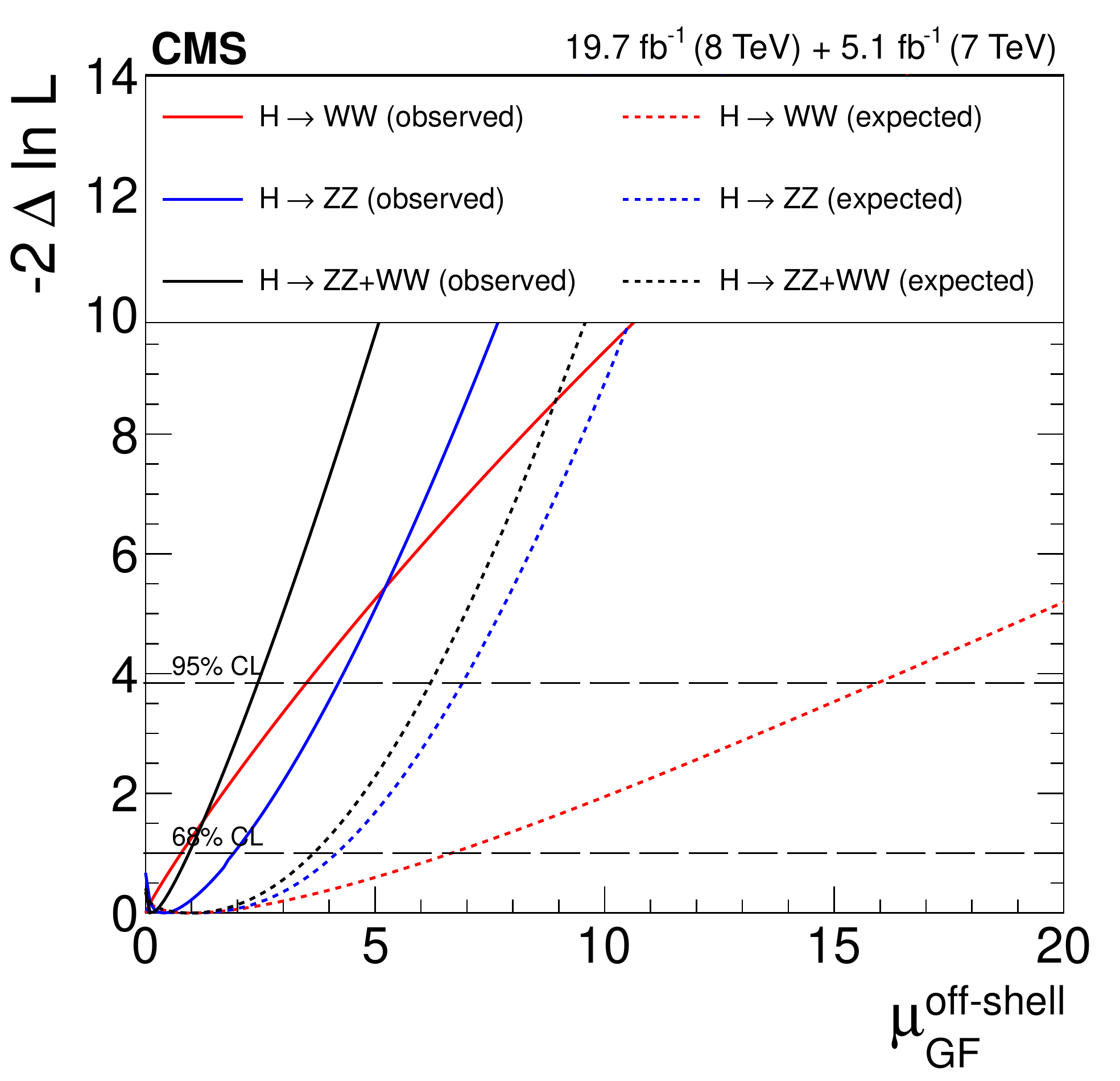}
     \includegraphics[width=0.48\textwidth]{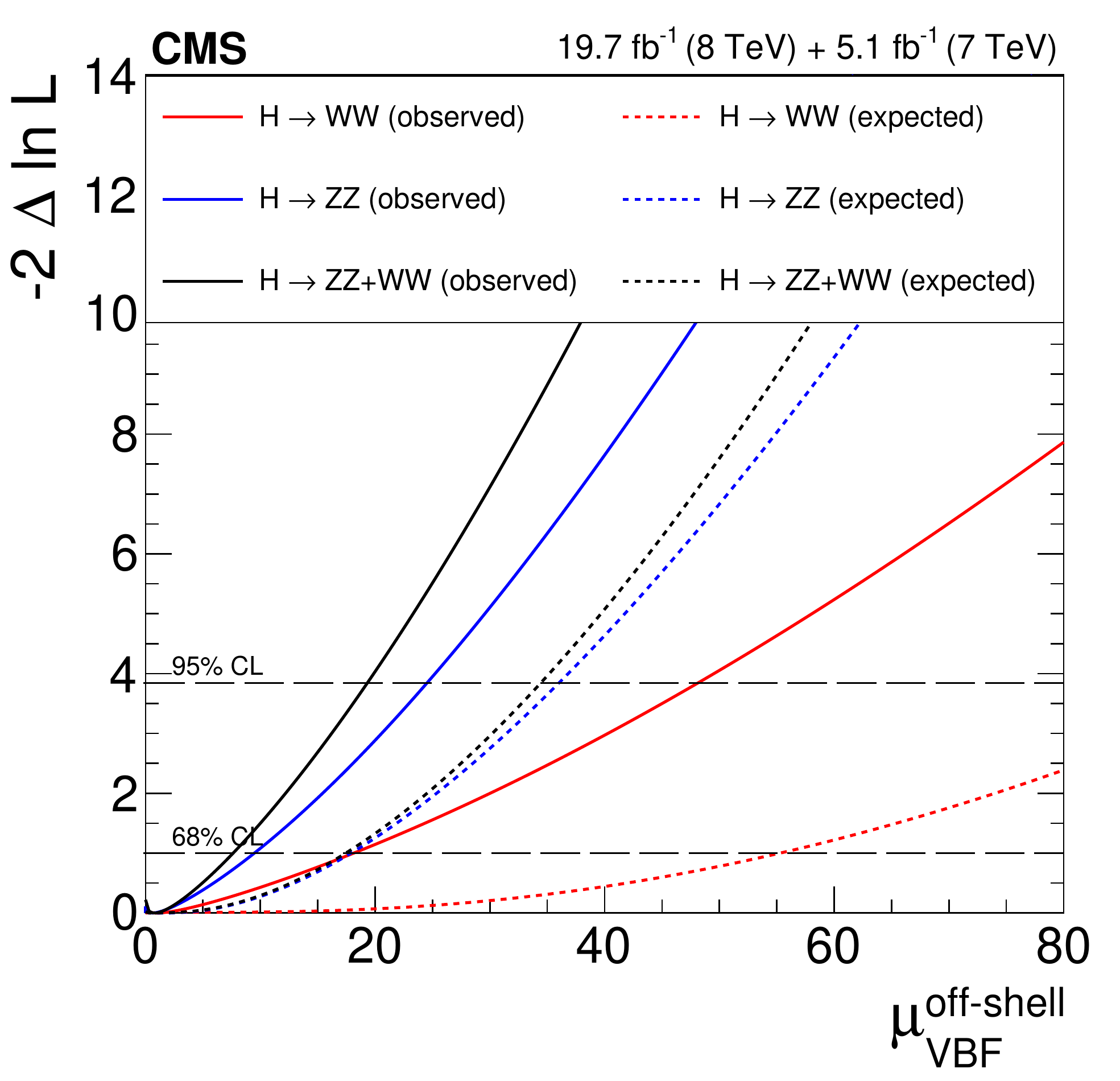}
     \caption{Scan of the negative log-likelihood as a function of off-shell SM Higgs boson signal strength for GF $\HstrOffGF$ (left) and
              for VBF $\HstrOffVBF$ (right)
	      from the combined fit of {$\PH\to\PW\PW$} and {$\PH\to\cPZ\cPZ$} channels for 7 and 8\TeV.
	      In the likelihood scan of $\HstrOffGF$ and $\HstrOffVBF$, this analysis assumes the SU(2) custodial symmetry: $\HstrZzGF/\HstrWwGF = \HstrZzVBF/\HstrWwVBF = 1$.}
\label{fig:combinationStrength}
\end{figure*}

The likelihood of the off-shell signal strength is scanned with the assumption of SU(2) custodial symmetry for the combination: $\HstrZzGF/\HstrWwGF = \HstrZzVBF/\HstrWwVBF= \Lambda_{\PW\cPZ}=1$.
The observed (expected) constraints on the off-shell signal strengths at 95\% CL are {$\HstrOffGF <2.4\,(6.2)$}
and {$\HstrOffVBF <19.3\,(34.4)$}, as shown in Fig.~\ref{fig:combinationStrength}.

\begin{figure*}
\centering
     \includegraphics[width=0.7\textwidth]{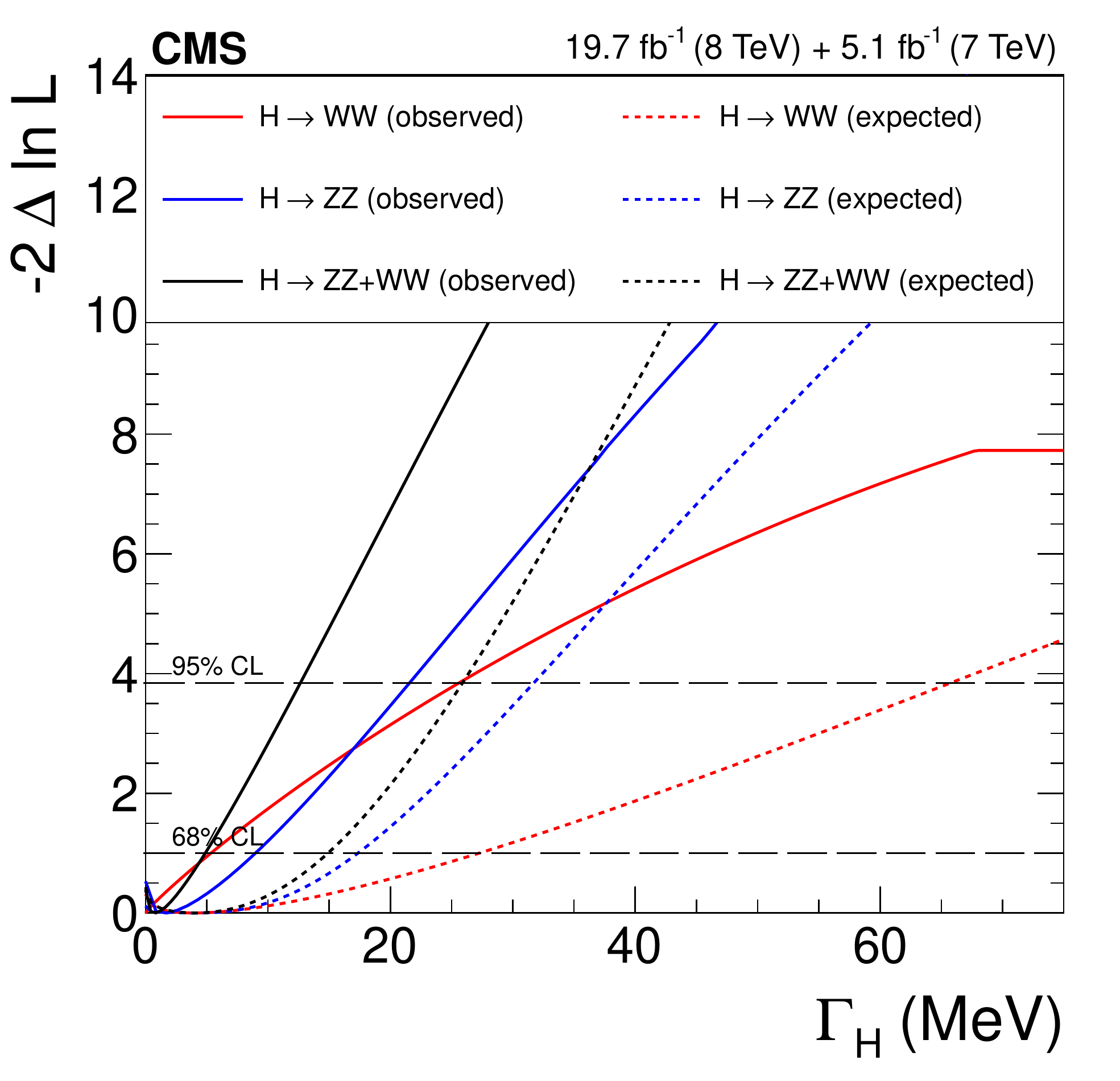}
     \caption{Scan of the negative log-likelihood as a function of {$\GH$}
              from the combined fit of {$\PH\to\PW\PW$} and {$\PH\to\cPZ\cPZ$} channels for 7 and 8\TeV.
	      In the likelihood scan of {$\GH$}, this analysis assumes the same GF and VBF ratio of signal strengths for $\PW\PW$ and $\cPZ\cPZ$ decay modes : $\HstrZzGF/\HstrWwGF = \HstrZzVBF/\HstrWwVBF$.
}
\label{fig:combination}
\end{figure*}

For the likelihood scan of $\GH$,
this analysis considers the possible difference of signal strength measurements between the two Higgs boson decay modes with an assumption
that the ratio of signal strengths is the same for each GF and VBF processes.
Accordingly, $\HstrWwGF$, $\HstrWwVBF$, $\HstrZzGF$, and $\HstrZzVBF$ can be expressed in terms of three independent parameters left floating in the fit:
$\HstrGF$, $\HstrVBF$, and $\Lambda_{\PW\cPZ}$:
$\HstrWwGF$ = $\HstrGF$, $\HstrWwVBF$ = $\HstrVBF$, $\HstrZzGF$ = $\Lambda_{\PW\cPZ}\HstrGF$, and $\HstrZzVBF$ = $\Lambda_{\PW\cPZ}\cdot \HstrVBF$,
where $\HstrGF$ and $\HstrVBF$ are the Higgs boson signal strengths for the GF and VBF production as in Eq.~(\ref{eq:pdf-prob-vbf})
and $\Lambda_{\PW\cPZ}$ is the common ratio $\HstrZzGF/\HstrWwGF = \HstrZzVBF/\HstrWwVBF = \Lambda_{\PW\cPZ}$.
Figure~\ref{fig:combination} shows the combined likelihood scan as a function of the Higgs boson width.
The observed (expected) combined limit for the width corresponds to 13 (26)\MeV at 95\% CL.
The observed limit improves by 50\% the result of the {$\PH\to\PW\PW$} channel alone ($<$26\MeV)
and by 41\% the observed limit of $<22$\MeV set in the {$\PH\to\cPZ\cPZ$} channel alone~\cite{CMS:2014ala}.
The result is about a factor of 3 larger than the SM expectation of $\GH \approx 4\MeV$.
Using pseudo data generated with the SM Higgs boson width,
the {$p$-value} for the observed limit is 7.4\%.
The relaxation of the same GF and VBF signal strength {$\cPZ\cPZ/\PW\PW$} ratios
increases the observed combined 95\% CL limit on the width to $\GH < 15\MeV$.

\section{Summary}

A search is presented for the Higgs boson off-shell production in gluon fusion and vector boson fusion processes with the Higgs boson decaying into a {$\WW$} pair and the {$\PW$} bosons decaying leptonically.
The data observed in this analysis are used to constrain the Higgs boson total decay width.
The analysis is based on {$\Pp\Pp$}~collision data collected by the CMS experiment at $\sqrt{s} = 7$ and 8\TeV, corresponding to integrated luminosities of 4.9 and 19.4\fbinv respectively.
The observed and expected upper limits for the off-shell signal strengths at 95\% CL are 3.5 and 16.0 for the gluon fusion process, and 48.1 and 99.2 for the vector boson fusion process.
The observed and expected constraints on the Higgs boson total width are, respectively, $\GH < 26$ and $<$66\MeV, obtained at the 95\% CL.
These results are combined with those obtained earlier in the {$\PH\to\cPZ\cPZ$} channel, which further improves the observed and expected upper limits of the off-shell signal strengths to 2.4 and 6.2 for the gluon fusion process, and 19.3 and 34.4 for the vector boson fusion process.
The observed and expected constraints on the Higgs boson total width from the combination are, respectively, $\GH < 13$ and $<$26\MeV at the 95\% CL.

\begin{acknowledgments}
\hyphenation{Bundes-ministerium Forschungs-gemeinschaft Forschungs-zentren} We congratulate our colleagues in the CERN accelerator departments for the excellent performance of the LHC and thank the technical and administrative staffs at CERN and at other CMS institutes for their contributions to the success of the CMS effort. In addition, we gratefully acknowledge the computing centres and personnel of the Worldwide LHC Computing Grid for delivering so effectively the computing infrastructure essential to our analyses. Finally, we acknowledge the enduring support for the construction and operation of the LHC and the CMS detector provided by the following funding agencies: the Austrian Federal Ministry of Science, Research and Economy and the Austrian Science Fund; the Belgian Fonds de la Recherche Scientifique, and Fonds voor Wetenschappelijk Onderzoek; the Brazilian Funding Agencies (CNPq, CAPES, FAPERJ, and FAPESP); the Bulgarian Ministry of Education and Science; CERN; the Chinese Academy of Sciences, Ministry of Science and Technology, and National Natural Science Foundation of China; the Colombian Funding Agency (COLCIENCIAS); the Croatian Ministry of Science, Education and Sport, and the Croatian Science Foundation; the Research Promotion Foundation, Cyprus; the Ministry of Education and Research, Estonian Research Council via IUT23-4 and IUT23-6 and European Regional Development Fund, Estonia; the Academy of Finland, Finnish Ministry of Education and Culture, and Helsinki Institute of Physics; the Institut National de Physique Nucl\'eaire et de Physique des Particules~/~CNRS, and Commissariat \`a l'\'Energie Atomique et aux \'Energies Alternatives~/~CEA, France; the Bundesministerium f\"ur Bildung und Forschung, Deutsche Forschungsgemeinschaft, and Helmholtz-Gemeinschaft Deutscher Forschungszentren, Germany; the General Secretariat for Research and Technology, Greece; the National Scientific Research Foundation, and National Innovation Office, Hungary; the Department of Atomic Energy and the Department of Science and Technology, India; the Institute for Studies in Theoretical Physics and Mathematics, Iran; the Science Foundation, Ireland; the Istituto Nazionale di Fisica Nucleare, Italy; the Ministry of Science, ICT and Future Planning, and National Research Foundation (NRF), Republic of Korea; the Lithuanian Academy of Sciences; the Ministry of Education, and University of Malaya (Malaysia); the Mexican Funding Agencies (BUAP, CINVESTAV, CONACYT, LNS, SEP, and UASLP-FAI); the Ministry of Business, Innovation and Employment, New Zealand; the Pakistan Atomic Energy Commission; the Ministry of Science and Higher Education and the National Science Centre, Poland; the Funda\c{c}\~ao para a Ci\^encia e a Tecnologia, Portugal; JINR, Dubna; the Ministry of Education and Science of the Russian Federation, the Federal Agency of Atomic Energy of the Russian Federation, Russian Academy of Sciences, and the Russian Foundation for Basic Research; the Ministry of Education, Science and Technological Development of Serbia; the Secretar\'{\i}a de Estado de Investigaci\'on, Desarrollo e Innovaci\'on and Programa Consolider-Ingenio 2010, Spain; the Swiss Funding Agencies (ETH Board, ETH Zurich, PSI, SNF, UniZH, Canton Zurich, and SER); the Ministry of Science and Technology, Taipei; the Thailand Center of Excellence in Physics, the Institute for the Promotion of Teaching Science and Technology of Thailand, Special Task Force for Activating Research and the National Science and Technology Development Agency of Thailand; the Scientific and Technical Research Council of Turkey, and Turkish Atomic Energy Authority; the National Academy of Sciences of Ukraine, and State Fund for Fundamental Researches, Ukraine; the Science and Technology Facilities Council, UK; the US Department of Energy, and the US National Science Foundation.

Individuals have received support from the Marie-Curie programme and the European Research Council and EPLANET (European Union); the Leventis Foundation; the A. P. Sloan Foundation; the Alexander von Humboldt Foundation; the Belgian Federal Science Policy Office; the Fonds pour la Formation \`a la Recherche dans l'Industrie et dans l'Agriculture (FRIA-Belgium); the Agentschap voor Innovatie door Wetenschap en Technologie (IWT-Belgium); the Ministry of Education, Youth and Sports (MEYS) of the Czech Republic; the Council of Science and Industrial Research, India; the HOMING PLUS programme of the Foundation for Polish Science, cofinanced from European Union, Regional Development Fund; the Mobility Plus programme of the Ministry of Science and Higher Education (Poland); the OPUS programme of the National Science Center (Poland); MIUR project 20108T4XTM (Italy); the Thalis and Aristeia programmes cofinanced by EU-ESF and the Greek NSRF; the National Priorities Research Program by Qatar National Research Fund; the Programa Clar\'in-COFUND del Principado de Asturias; the Rachadapisek Sompot Fund for Postdoctoral Fellowship, Chulalongkorn University (Thailand); the Chulalongkorn Academic into Its 2nd Century Project Advancement Project (Thailand); and the Welch Foundation, contract C-1845.

\end{acknowledgments}
\clearpage
\bibliography{auto_generated}

\cleardoublepage \appendix\section{The CMS Collaboration \label{app:collab}}\begin{sloppypar}\hyphenpenalty=5000\widowpenalty=500\clubpenalty=5000\textbf{Yerevan Physics Institute,  Yerevan,  Armenia}\\*[0pt]
V.~Khachatryan, A.M.~Sirunyan, A.~Tumasyan
\vskip\cmsinstskip
\textbf{Institut f\"{u}r Hochenergiephysik der OeAW,  Wien,  Austria}\\*[0pt]
W.~Adam, E.~Asilar, T.~Bergauer, J.~Brandstetter, E.~Brondolin, M.~Dragicevic, J.~Er\"{o}, M.~Flechl, M.~Friedl, R.~Fr\"{u}hwirth\cmsAuthorMark{1}, V.M.~Ghete, C.~Hartl, N.~H\"{o}rmann, J.~Hrubec, M.~Jeitler\cmsAuthorMark{1}, A.~K\"{o}nig, M.~Krammer\cmsAuthorMark{1}, I.~Kr\"{a}tschmer, D.~Liko, T.~Matsushita, I.~Mikulec, D.~Rabady, N.~Rad, B.~Rahbaran, H.~Rohringer, J.~Schieck\cmsAuthorMark{1}, R.~Sch\"{o}fbeck, J.~Strauss, W.~Treberer-Treberspurg, W.~Waltenberger, C.-E.~Wulz\cmsAuthorMark{1}
% ...
\vskip\cmsinstskip
% ...
1:~~Also at Vienna University of Technology, Vienna, Austria\\
% ...

\end{sloppypar}
\end{document}